\documentclass[preprint,showpacs,nofootinbib]{revtex4}
\usepackage{color}
\usepackage{grffile}
\usepackage{graphicx}
\newcommand{\ber}{\begin{eqnarray}}
\newcommand{\eer}{\end{eqnarray}}
\newcommand{\fqgp}{f_{\rm QGP}}
\newcommand{\fqgpb}{f_{\rm QGP,B}}
\newcommand{\gb}{g_{\rm B}}
\newcommand{\vtwoep}{v_2\{\rm EP\}}
\newcommand{\pt}{p_{\rm T}} 
\newcommand{\pz}{p_{\rm z}} 
\newcommand{\epcm}{\epsilon^{\rm c.m.}} 
\newcommand{\te}{T_\epsilon} 
\newcommand{\tet}{T_{\epsilon_{\rm T}}}
\newcommand{\tecm}{T_{\epsilon}^{\rm c.m.}}
\newcommand{\tn}{T_n} 

\newcommand{\tmet}{T_{\left < E_{\rm T} \right >}}
\newcommand{\tmpt}{T_{\left < p_{\rm T} \right >}} 
\newcommand{\tmptsq}{T_{\left < \pt^2 \right >}}
\newcommand{\tmp}{T_{\left < p \right >}} 

\begin{document}
\title{Evolution of transverse flow and effective temperatures in the parton phase 
from a multi-phase transport model}
\author{Zi-Wei Lin\footnote{linz@ecu.edu}}
\affiliation{Department of Physics, East Carolina University,
  C-209 Howell Science Complex, Greenville, NC 27858}

\begin{abstract}
I study the space-time evolution of transverse flow and effective
temperatures in the dense parton phase with the string melting
version of a multi-phase transport model. 
Parameters of the model are first constrained to reproduce the bulk 
data on the rapidity density, $\pt$ spectrum and elliptic flow at low
$\pt$ for central and mid-central Au+Au collisions at $200A$ GeV and
Pb+Pb collisions at $2760A$ GeV. I then calculate the transverse flow
and effective temperatures in volume cells within
mid-spacetime-rapidity $|\eta|<1/2$. 
I find that the effective temperatures extracted from different
variables, which are all evaluated in the rest frame of a volume cell,
can be very different; this indicates that the parton system in the model
is not in full chemical or thermal equilibrium locally, even after
averaging over many events. 
In particular, the effective temperatures extracted from the parton
energy density or number density  are often quite different than those
extracted from the parton mean $\pt$ or mean energy. 
For these collisions in general, effective temperatures extracted from
the parton energy density or number density are higher than those
extracted from the parton mean $\pt$ in the inner part of the overlap
volume, while the opposite occurs in the outer part of the overlap
volume. 
I argue that this indicates that the dense parton matter in the inner
part of the overlap volume is over-populated; I also find that all cells
with energy density above 1 GeV/fm$^3$ are over-populated after a
couple of fm/$c$.  

\end{abstract}
\maketitle

\section{Introduction}

A dense matter consisting of partonic degrees of freedom, often called
the quark-gluon plasma (QGP), has been created in ultra-relativistic heavy 
ion collisions at the Relativistic Heavy Ion Collider (RHIC) and the
Large Hadron Collider (LHC).
Its properties such as the color glass condensate initial condition, 
the degree of equilibration, event-by-event fluctuations, and coupling
with high-momentum partons are being extensively studied
\cite{Alver:2010gr,Blaizot:2011xf,CMS:2012qk,Betz:2012qq}.
Simulations of these collisions with hydrodynamic codes
\cite{Huovinen:2001cy,Betz:2008ka,Schenke:2010rr,Bozek:2011if}, transport models
\cite{Xu:2004mz,Lin:2004en,Cassing:2009vt}, or hybrid models
\cite{Petersen:2008dd,Werner:2010aa,Song:2010mg} are able to produce
the full evolution history of dense matter and are very useful for the
studies of the quark-gluon plasma properties. 
In this paper I investigate the evolution of the parton matter at
mid-spacetime-rapidity, including its transverse flow and effective
temperatures. The string melting version of the AMPT model
\cite{Lin:2004en} is used here, where
excited hadronic strings in the overlap volume are converted into
partons via the intermediate step of decomposing hadrons that would
have been produced by the Lund string fragmentation process
\cite{Lin:2001zk,Lin:2004en}. 

The default version of the AMPT model \cite{Zhang:1999bd,Lin:2000cx}
was first constructed to simulate relativistic heavy ion collisions,
and its key parameters were determined to fit the yields and
$\pt$ spectra of particles in pp collisions at various energies and 
heavy ion collisions up to SPS energies. 
Hadronization in the default AMPT model is described by the Lund
string model \cite{Sjostrand:1993yb}, where one assumes that a string
fragments into quark-antiquark pairs with a Gaussian distribution in
transverse momentum. 
Hadrons are formed from these quarks and antiquarks, with its
longitudinal momentum given by the Lund symmetric fragmentation
function $f(z) \propto z^{-1} (1-z)^a \exp (-b~m_{\rm T}^2/z)$, 
where $z$ represents the light-cone momentum fraction of the produced
hadron with respect to that of the fragmenting string and $m_{\rm T}$
is the transverse mass of the hadron. 
It was found \cite{Zhang:1999bd,Lin:2000cx} 
that the default HIJING values for the Lund string
fragmentation parameters ($a=0.5$ and $b=0.9$ GeV$^{-2}$), which work
well for pp collisions, led to too small a charged particle yield in 
central Pb+Pb collisions at the SPS energy of $E_{\rm
  LAB}=158A$ GeV. Therefore modified values of $a=2.2$ and $b=0.5$
GeV$^{-2}$ were used in order to fit the charged particle yield
in Pb+Pb collisions at SPS.   
For heavy ion collisions at higher energies such as RHIC energies, 
the default version of the AMPT model with these parameters values 
was found to reasonably fit dN/d$\eta$, dN/dy and the $\pt$ spectra in
heavy ion collisions, although it under-estimates the elliptic flow. 
On the other hand, the string melting version of the AMPT model
(AMPT-SM) \cite{Lin:2001zk,Lin:2004en}, due to its dense parton phase,
reasonably fits the elliptic flow and two-pion HBT in heavy ion
collisions; but it does not reproduce well 
dN/d$\eta$, dN/dy and the $\pt$ spectra (when using the same
parameters as in the default version). In particular, the AMPT-SM model
significantly over-estimates the charged particle yield while 
under-estimates the slopes of the $\pt$ spectra \cite{Lin:2004en}. 
In an earlier attempt to reproduce data in Pb+Pb collisions at LHC
energies with the AMPT-SM model, the default HIJING values for the
Lund string fragmentation parameters were used \cite{Xu:2011fi}
together with the strong coupling  constant $\alpha_s=0.33$ (instead
of $0.47$); there the model reasonably reproduced the yield and
elliptic flow of charged particles but underestimated the $\pt$
spectrum (except at low $\pt$). 

In this study I use the AMPT-SM model to study the evolution of the
dense parton phase. I first tune the key parameters of the model
to reproduce the pion and kaon yields, $\pt$ spectra, 
and elliptic flows at low $\pt$ (below $\sim 2$ GeV/$c$) in central
and mid-central Au+Au collisions at the RHIC energy of $200A$ GeV and
Pb+Pb collisions at the LHC energy of $2760A$ GeV.
I then study the evolution of transverse flow and effective
temperature in the parton phase. 
In the analysis, the reaction volume within $|\eta|<1/2$ is divided 
into cells with a transverse width of $1$ fm.  
In order to have enough statistics for the analysis of each volume
cell, I choose to study event-averaged quantities and thus 
neglect event-by-event fluctuations in this study. 
For each collision energy and centrality, I run hundreds to
thousands of events at the same impact parameter range and then
calculate event-averaged quantities at functions of time and
transverse location. 
Note that the AMPT-SM model assigns a formation time
to each parton produced from string melting \cite{Lin:2004en}, and
only partons that have formed by a global time $t$ are included in the
analysis for that time. 
The full space-time evolutions of the parton matter from this study
have been posted online \footnote{Grid data files of the space-time
  evolution of the parton matter from this study have been posted 
\cite{datalink}; the link has also been posted at the JET
  Collaboration wiki page \cite{jetwiki}.}  
and may be used as the bulk matter 
background within which other probes such as jet propagation and
interactions can be studied \cite{Betz:2013caa,Burke:2013yra}.

\section{Fitting the bulk data with the string melting version of AMPT}

Since the purpose of the study is to obtain the space-time evolution
of the dense parton matter, I first constrain the model
parameters by fitting low $\pt$ data in heavy ion collisions at RHIC
and LHC energies. To fit the low $\pt$ data in central and mid-central
Au+Au collisions at the top RHIC energy,  I find that I need to set
the values of the Lund string fragmentation parameters to $a=0.55$,
$b=0.15$ GeV$^{-2}$. 
Note that I use a lower $b$ value than in previous studies 
\cite{Zhang:1999bd,Lin:2000cx,Lin:2001zk,Lin:2004en,Xu:2011fi}
in order to simultaneously fit the rapidity density, $\pt$ spectrum
and elliptic flow of pions and kaons at low $\pt$ with the
string melting version of the AMPT model.  
Also note that the effective string tension, as given by
$\kappa \propto 1/[b(2+a)]$ \cite{Lin:2004en}, will change when the
Lund string fragmentation parameters are modified; and a smaller $b$
value will lead to a larger mean transverse momentum of the initial hadrons. 
In addition, the AMPT model assumes that the relative production of
strange to non-strange quarks increases with the effective string
tension \cite{Lin:2004en}.  
Since the above string tension relation diverges as $b \rightarrow 0$ 
and this study uses a small $b$ value to enable the AMPT-SM results to 
fit the data, I put an upper limit of 0.40 on the relative production
of strange to non-strange quarks in AMPT. For Pb+Pb collisions at the
LHC energy, I use $a=0.30$  (and $b=0.15$ GeV$^{-2}$) in order to fit
the ALICE data \cite{Abelev:2013vea}.    
I take $\alpha_s=0.33$ and a parton cross section of $3$ mb  
for all simulations in this study.
There may be other AMPT parameter sets that could reasonably fit the
data. For example, I find that the Lund $a$ value of $0.55$ used for
fitting the RHIC data gives a reasonable but less 
satisfactory fit of the ALICE data than the $a$ value of $0.30$. 
One may also use a lower value of the strong coupling $\alpha_s$ at
LHC than at RHIC \cite{Betz:2012qq} and tune other AMPT parameters to
obtain such a simultaneous fit to data.

\begin{figure}[h]
\includegraphics[width=6 in]{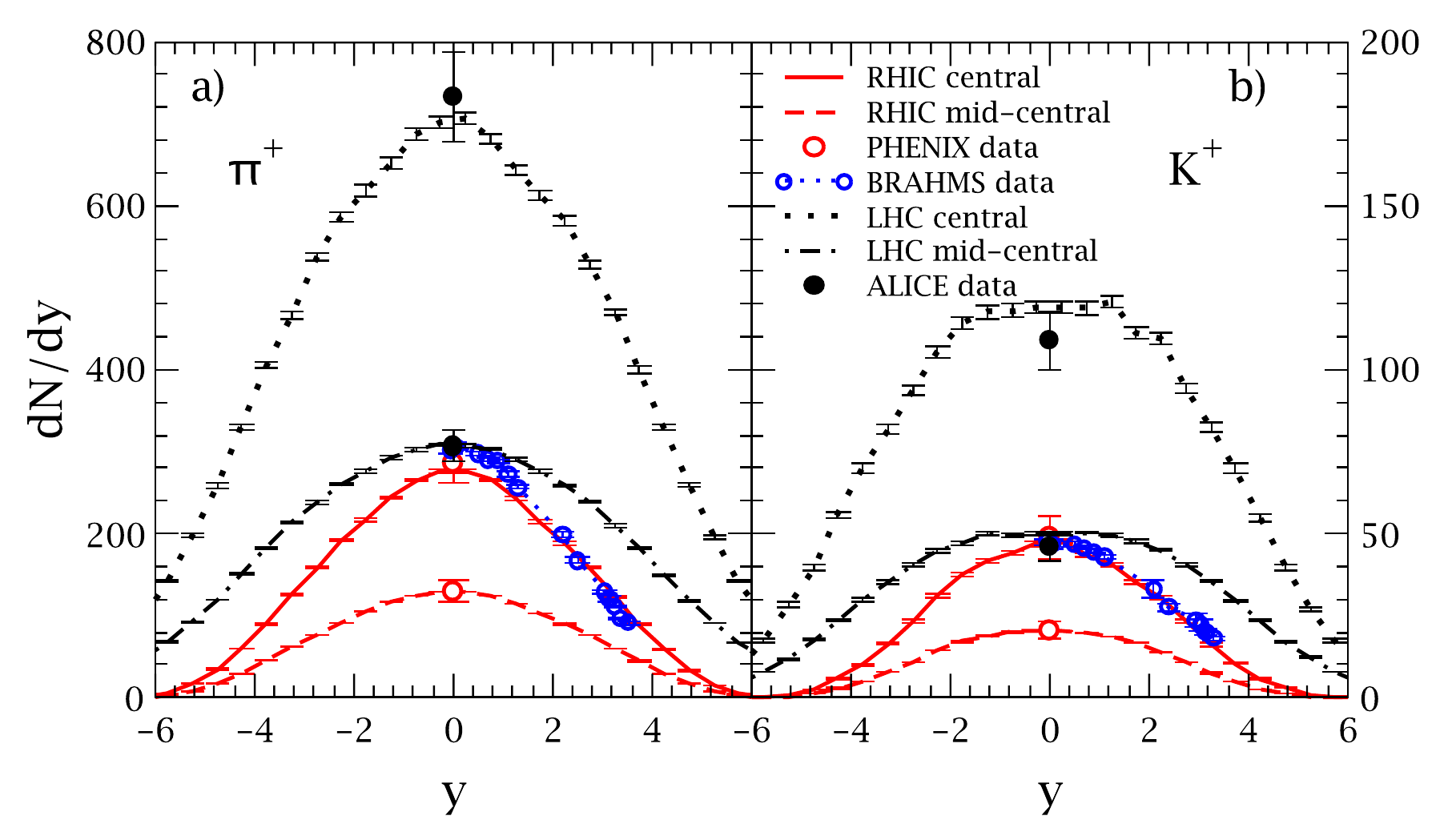}
\caption{(Color online) AMPT-SM results for central and mid-central Au+Au collisions
  at $200A$ GeV and Pb+Pb collisions at $2760A$ GeV in comparison with
  experimental data for 0-5\% and 20-30\% centralities: 
a) dN/dy of $\pi^+$,
and b) dN/dy of $K^+$.
}
\label{fig:dndy}
\end{figure}
 
\begin{figure}[h]
\includegraphics[width=6 in]{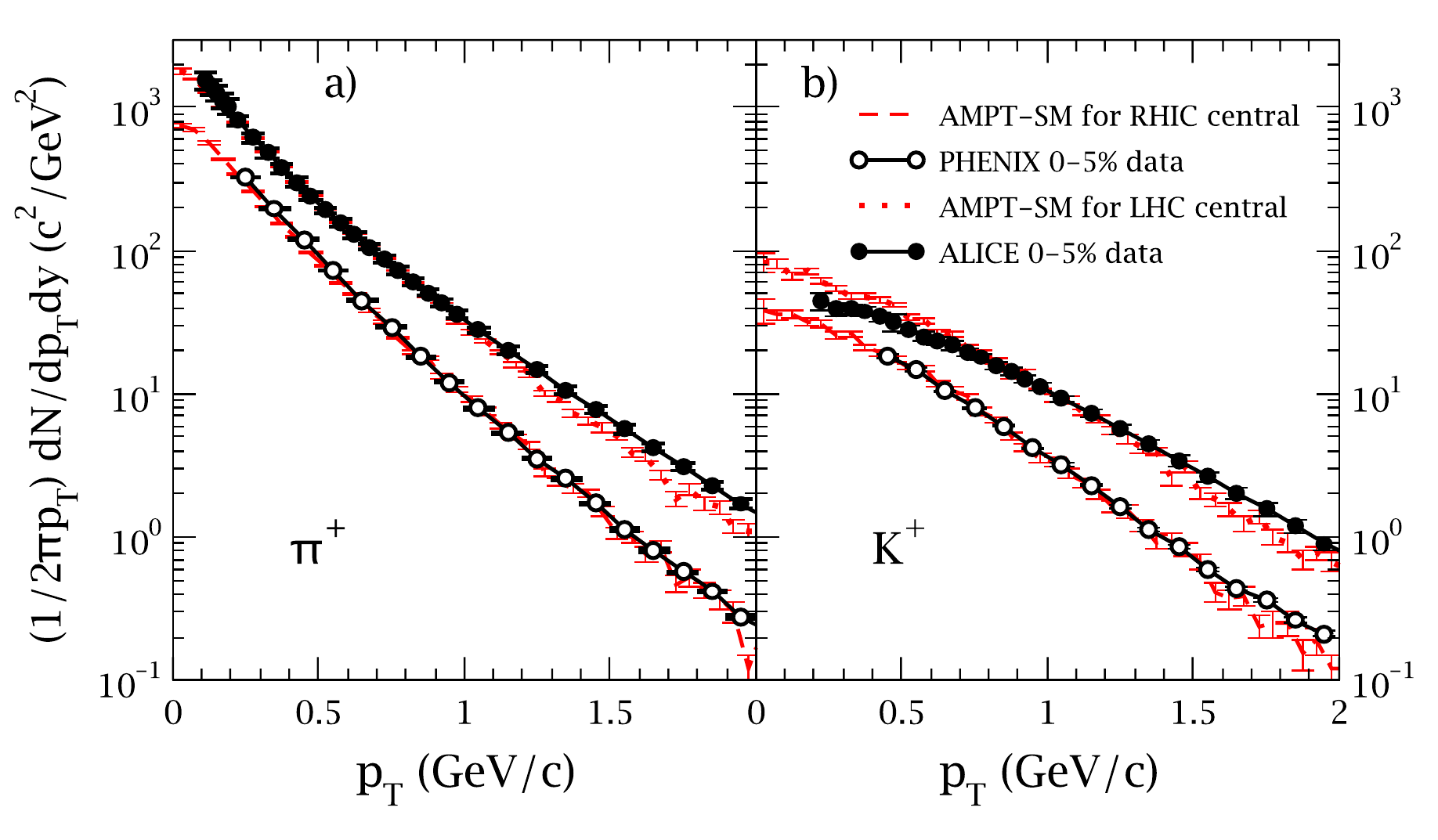}
\caption{(Color online) 
$\pt$ spectra of $\pi^+$ and $K^+$ at mid-rapidity in central
collisions from AMPT-SM in comparison with experimental data for 0-5\%
centrality.
}
\label{fig:pt}
\end{figure}

The AMPT-SM results shown in this study include four collision systems: 
RHIC central refers to Au+Au events at $200A$ GeV with $b<3$ fm 
that represent the 0-5\% centrality,  
RHIC mid-central refers to Au+Au events at $200A$ GeV with $b=7.3$ 
fm that represent the 20-30\% centrality \cite{Adare:2012wg},  
LHC central refers to Pb+Pb events at $2760A$ GeV 
with $b<3.5$ fm that represent the 0-5\% centrality,  
and LHC mid-central refers to Pb+Pb events at $2760A$ GeV with
$b=7.8$ fm that represent the 20-30\% centrality
\cite{Abelev:2013qoq}. 
Since each volume cell within $|\eta|<1/2$ has a transverse width of
$1.0$ fm along both the $x$- and $y$-axis, the cell at $x=3$~fm \&
$y=0$~fm then refers to the volume within $2.5<x<3.5$~fm
\& $-0.5<y<0.5$~fm \& $|\eta|<1/2$, for example.

The comparisons between AMPT-SM results and the experimental data on
particle dN/dy are shown in Fig.~\ref{fig:dndy}a  
for $\pi^+$ and in Fig.~\ref{fig:dndy}b for $K^+$, where we see
good agreements between the model results and the PHENIX
\cite{Adler:2003cb} and ALICE data \cite{Abelev:2013vea} 
at mid-rapidity in both central and mid-central events at RHIC and LHC
energies. Reasonable agreements are also seen in comparison with the
BRAHMS \cite{Bearden:2004yx} data at different rapidities 
in 0-5\% central Au+Au collisions at $200A$ GeV. 
The comparisons of the $\pi^+$ and $K^+$ $\pt$ spectra 
at mid-rapidity are shown in Fig.~\ref{fig:pt}a and 
Fig.~\ref{fig:pt}b respectively 
for central Au+Au collisions at the RHIC energy and central Pb+Pb
collisions at the LHC energy, where the model roughly reproduces
the observed $\pt$ spectra below 2 GeV/$c$.

\begin{figure}[h]
\includegraphics[width=6 in]{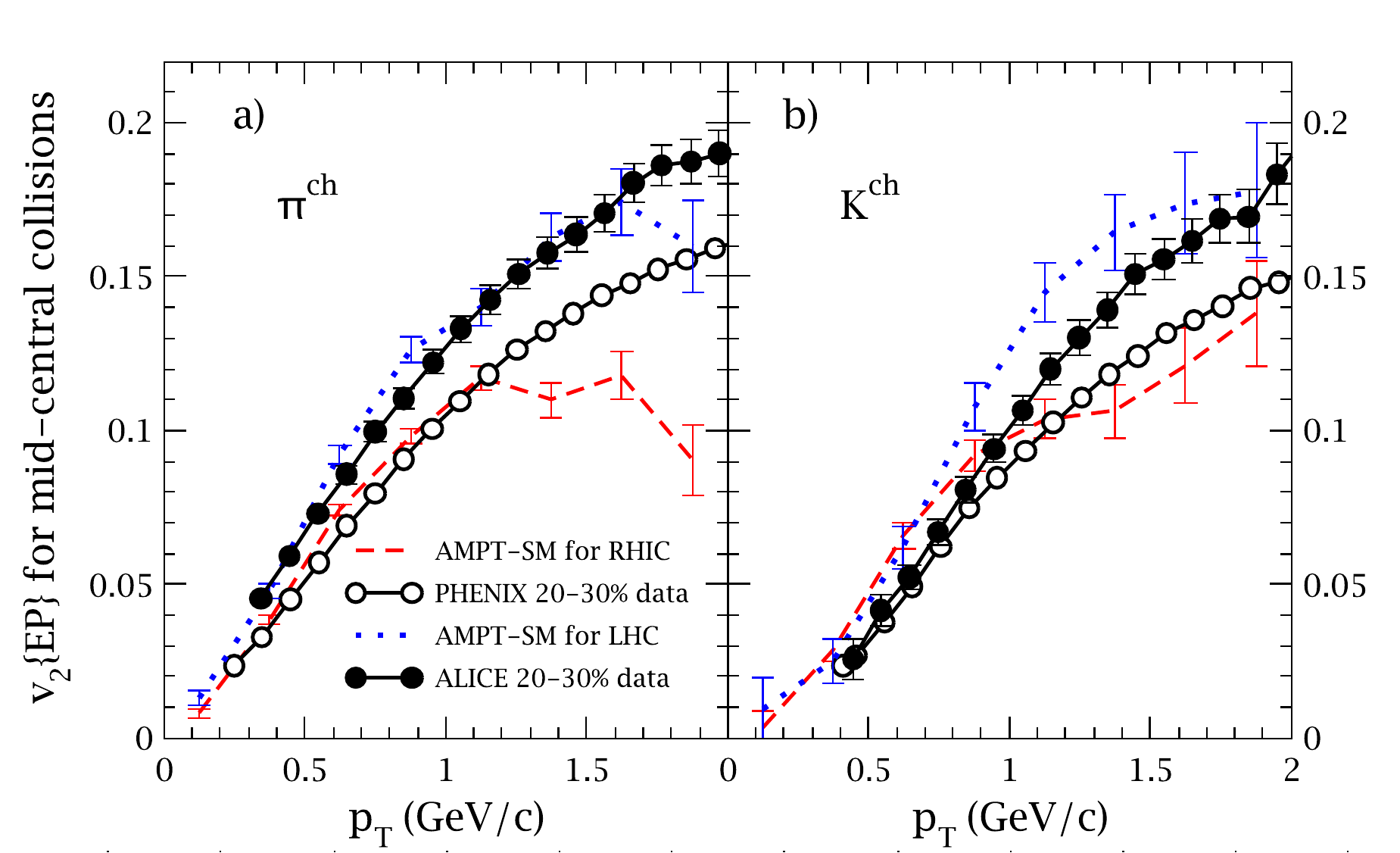}
\caption{(Color online) Elliptic flow $\vtwoep$ at mid-rapidity 
  in mid-central collisions from AMPT-SM (curves without circles) in comparison
  with experimental data for 20-30\% centrality (circles):
a) for charged pions,
and b) for charged kaons.
}
\label{fig:v2}
\end{figure}

Fig.~\ref{fig:v2} shows the comparisons of $\vtwoep$ of charged
pions and kaons. We see that the AMPT-SM model is also able to roughly
reproduce the pion and kaon elliptic flow data on $\vtwoep$
\cite{Gu:2012br} at low-$\pt$. 
Note that the correction factor for the event plane resolution
${\rm Res}\{2\Phi_2\}$ \cite{ATLAS:2012at} in the AMPT-SM results is 
calculated using particles within rapidities $-2.8<y<-1$ and within
$1<y<2.8$.

\section{Space-time evolution of transverse flow}

\begin{figure}[h]
\includegraphics[width=4 in]{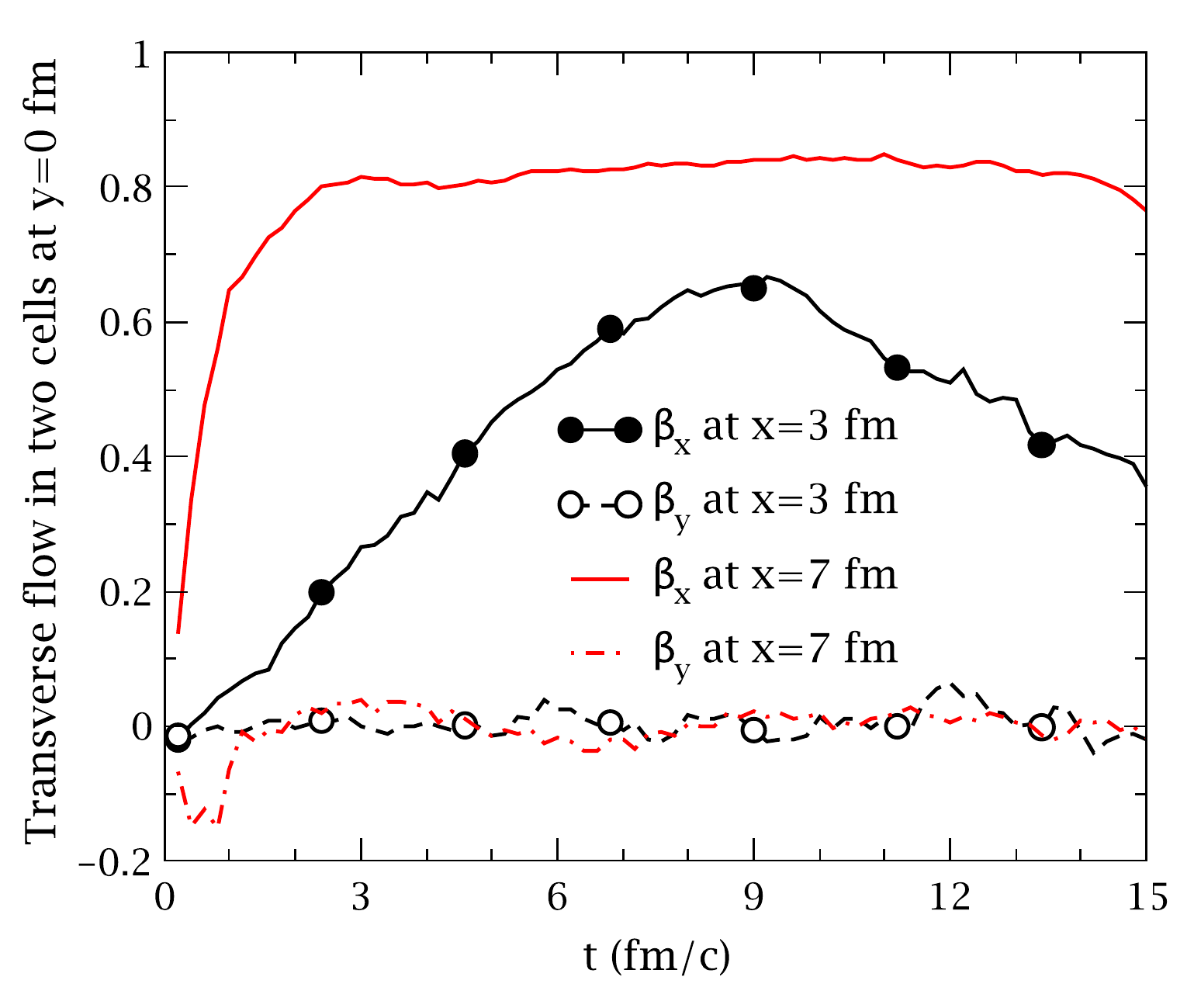}
\caption{(Color online)  
Transverse flow of partons in the cell at $x=3$~fm \& $y=0$~fm and in 
the cell at $x=7$~fm \& $y=0$~fm as functions of time in central Au+Au
collisions at $200A$ GeV.
}
\label{fig:flowt}
\end{figure}

For each volume cell within $|\eta|<1/2$, I calculate the flow as
$\vec \beta = \left (\sum_i \vec p_i \right )/\left (\sum_i E_i \right )$, 
where the sum over index $i$  takes into account all formed partons in
the cell from all events of a given collision system. 
Fig.~\ref{fig:flowt} shows the time evolutions of
the transverse flow of partons in two cells in central Au+Au
collisions at $200A$ GeV: 
one cell at $x=3$~fm \& $y=0$~fm and the other cell 
at $x=7$~fm \& $y=0$~fm. 
The flows along the $y$-direction in both 
cells here are essentially zero due to the symmetry
after averaging over many events.
For the flow along the $x$-direction, however, 
we see that both the profile and magnitude depend significantly on the
location even for these central collisions: 
the flow in the cell closer to the center of the overlap volume 
develops gradually and then decreases at later times, 
while the flow in the cell farther away from the center develops very
fast and reaches a bigger magnitude. 
Note that in the AMPT model the centers of the two incoming nuclei
are essentially at $x=b/2$ \& $y=0$ fm and $x=-b/2$ \& $y=0$ fm,
respectively. As a result, the center of the overlap region of
each Au+Au or Pb+Pb event is at $x=0$ fm \& $y=0$ fm, when
event-by-event fluctuations are neglected or when the calculation
averages over many events at the same impact parameter.

\begin{figure}[h]
\includegraphics[width=4 in]{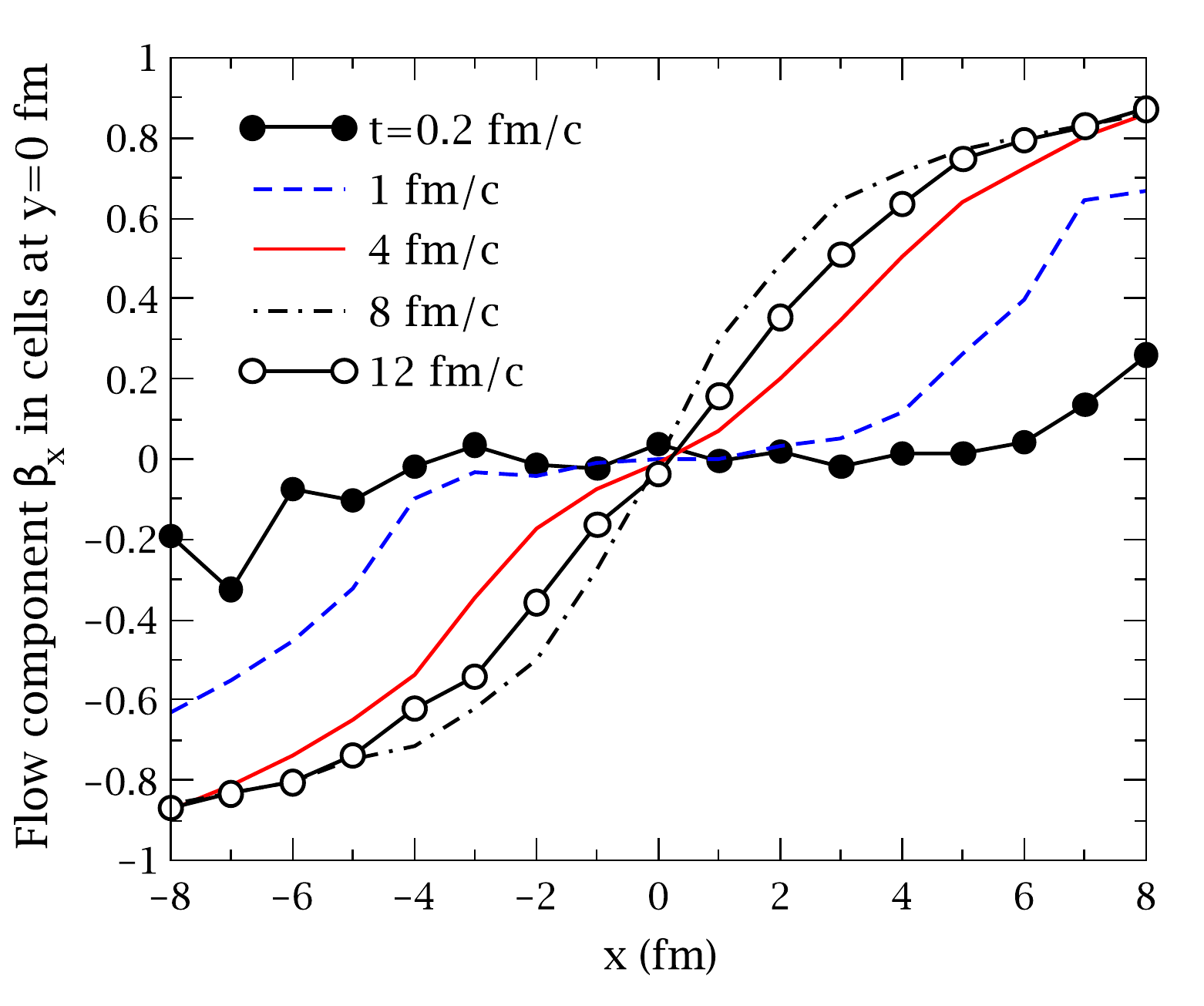}
\caption{(Color online)    
Flow along the $x$-direction as functions of $x$ at different times in 
cells within $-0.5<y<0.5$~fm in central Au+Au collisions at $200A$ GeV.
}
\label{fig:flowx}
\end{figure}
 
Fig.~\ref{fig:flowx} shows transverse flows along the $x$-direction 
in cells within $-0.5<y<0.5$~fm at different locations along
the $x$-axis at various times in central Au+Au collisions at RHIC. 
We see that the flow is initially very small at the early time $t=0.2$
fm/$c$ and then develops rather fast, while 
for the inner part of the overlap region ($-4<x<4$~fm) 
the flow decreases appreciably later in the evolution 
(after $t \sim 8$ fm/$c$). 
We also see that in general the flow magnitude is bigger the 
further away from the overlap center in these central collisions
regardless of time $t$ shown in the figure.

\section{Space-time evolution of temperature}

Partons in a given volume of a heavy ion collision, even after being
averaged over many events at the same impact parameter, may not be 
in full thermal equilibrium or full chemical equilibrium. 
If this is the case, the ``temperature'' for such a local parton
system will only be an effective temperature, which value depends on the
variable that it is extracted from. 
In this study I use different variables in the rest frame of a cell,
such as the parton mean $\pt$, number density and energy density, to
extract the effective temperature of partons in the cell.

To extract the effective temperature from the mean momentum of partons
in the rest frame of a cell,  I use the relations for a massless
parton gas in thermal equilibrium as given by the Boltzmann
distribution to get 
\ber
\tmp=\frac {\left < p \right >}{3}, 
\tmpt=\frac {4\left < \pt \right >}{3\pi}, 
\tmptsq=\sqrt {\frac {\left < \pt^2 \right >}{8}}.
\label{tpB}
\eer
In the above, the bracket represents the mean value of a variable, 
$p$ represents the magnitude of the parton 3-momentum, 
and $\pt$ represents the parton transverse momentum. 
Note that quark masses in the parton phase of the AMPT model are
current quark masses \cite{Lin:2004en}. Also, the parton phase in the
AMPT model does not include the effect of quantum statistics, and 
the Boltzmann distribution is assumed in the formulae and 
corresponding curves in this study unless specified otherwise.

I also extract the effective temperature from the number density or
energy density of partons in the rest frame of a cell. 
For this purpose I use the following relations between the densities
and temperature $T$ for a massless quark-gluon plasma in full chemical and 
thermal equilibrium as given by the Boltzmann distribution: 
\ber
n=\gb \frac {T^3}{\pi^2},
\epsilon_{\rm T}=3 \gb \frac {T^4}{4\pi},
\epsilon=3 \gb \frac {T^4}{\pi^2}.
\label{neB}
\eer
In the above, $n$ is the parton number density, 
$\epsilon_{\rm T}$ is the transverse energy density, 
$\epsilon$  is the energy density, 
$\gb \equiv 4(4+3N_f)$ is the total
degeneracy factor of QGP when using Boltzmann distributions, 
and $N_f$ represents the number of relevant quark flavors.  
I then have the following equations for the corresponding effective
temperatures:
\ber
\tn=168. {\rm MeV} \left ( \frac {n {~\rm fm^3}}{1+3N_f/4}
\right)^{1/3}, \nonumber \\
\tet=212. {\rm MeV} \left ( \frac {\epsilon_{\rm T} {~\rm
      fm^3/GeV}}{1+3N_f/4} \right)^{1/4}, \nonumber \\ 
\te=199. {\rm MeV} \left ( \frac {\epsilon {~\rm
      fm^3/GeV}}{1+3N_f/4} \right)^{1/4}.
\label{teB}
\eer
I use the above relations to extract effective
temperatures, and I take $N_f=3$ throughout this study.
Note that using Bose-Einstein and Fermi-Dirac
distributions for a massless quark-gluon plasma would lead to 
$n=(16+9N_f) \zeta(3) T^3/\pi^2$ and 
$\epsilon=\pi^2(16+10.5N_f) T^4/30$, 
where $\zeta(3) \simeq 1.20 $ is a Riemann zeta function. 
For $N_f=3$ they would give
$n \simeq 5.24 T^3$ and 
$\epsilon \simeq 15.6 T^4$, which 
are very close to the relations $n \simeq 5.27 T^3$ and $\epsilon
\simeq 15.8 T^4$ when the Boltzmann distribution is used for massless
partons as done in Eqs.~(\ref{neB}-\ref{teB}).  

\subsection{Time evolution of parton densities and mean momentum}

\begin{figure}[h]
\includegraphics[width=2.94 in]{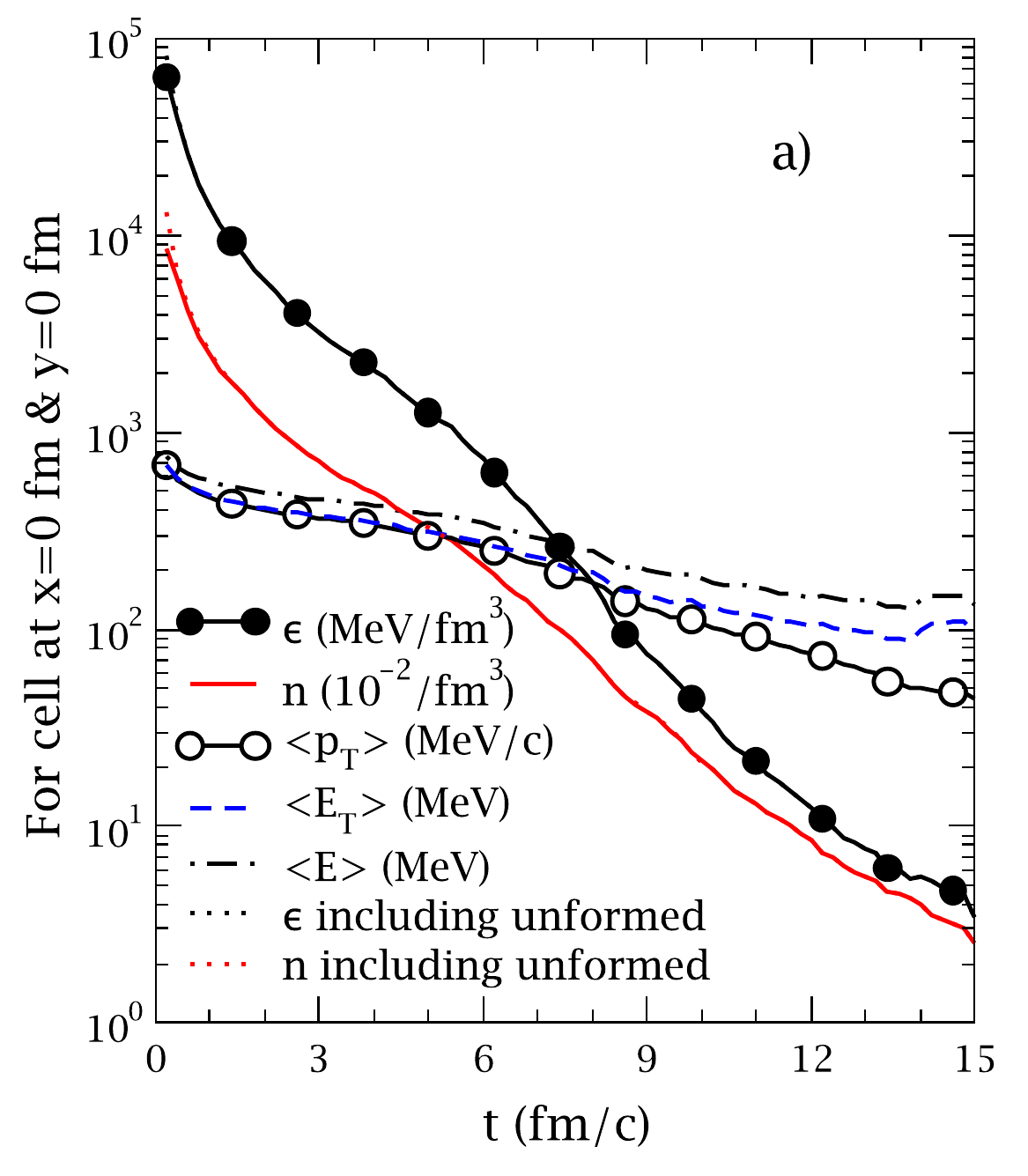}
\hspace{0.3cm}
\includegraphics[width=2.94 in]{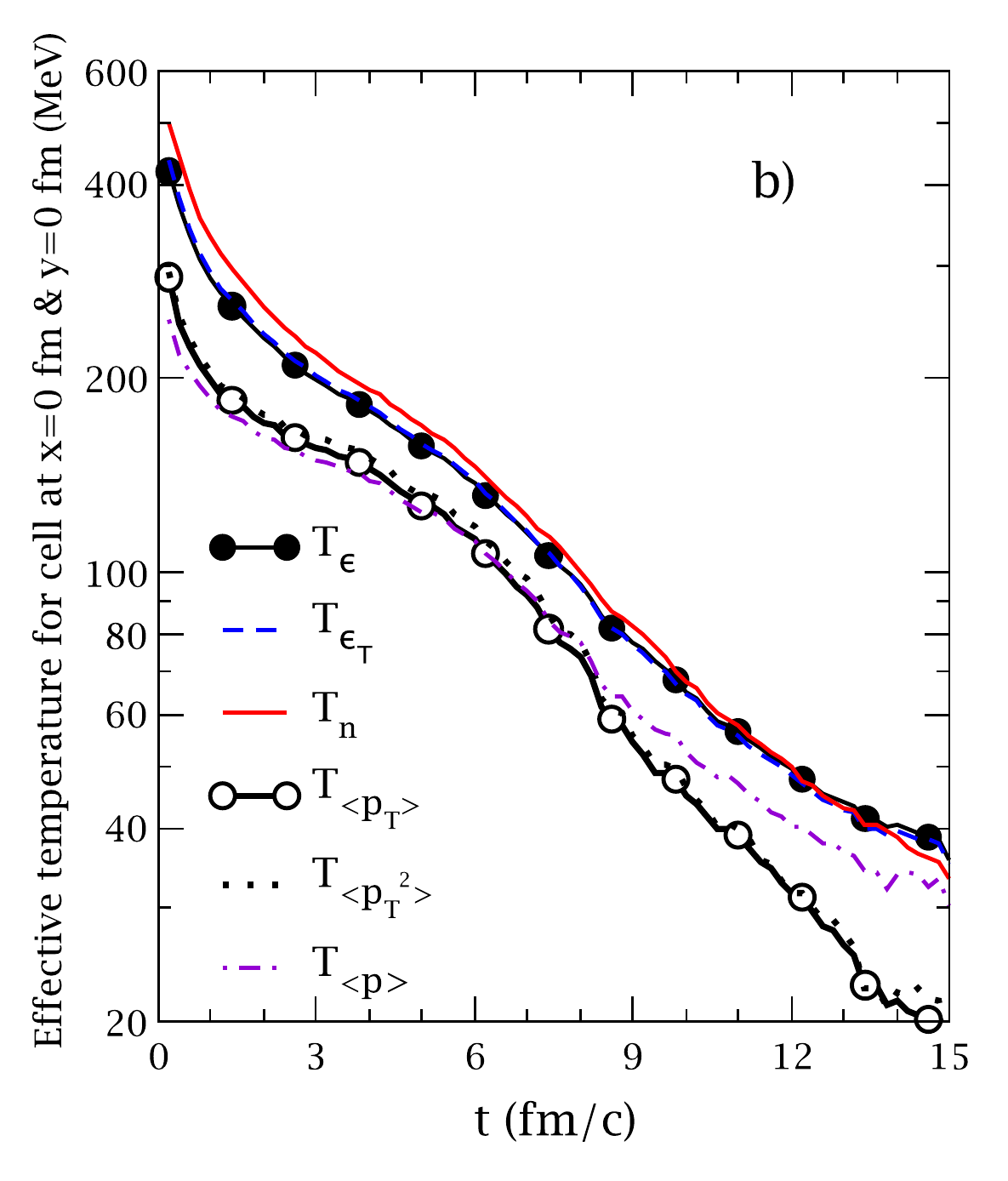}
\caption{(Color online) 
AMPT-SM results for partons in the rest frame of the center cell 
in central Au+Au collisions at $200A$ GeV:  
a) time evolutions of the energy density, number density, mean $\pt$, 
mean transverse energy, and mean energy; 
and b) time evolutions of effective temperatures extracted from
different variables.
}
\label{fig:center}
\end{figure}

I first examine the parton variables evaluated in the rest frame of
the cells, which I shall use to extract the effective temperatures. 
The left panel of Fig.~\ref{fig:center} shows the time evolutions of 
different variables in the center cell (at $x=0$~fm \&
$y=0$~fm) in central Au+Au collisions at $200A$ GeV;  
they include the parton energy density (filled circles), number
density (solid curve), mean $\pt$ (open circles), mean transverse
energy (dashed curve), and mean energy $\left < E \right >$
(dot-dashed curve).  The parton number density is seen to decrease
much faster with time than the parton mean $\pt$ or mean energy here. 
Consequently the parton energy density decreases even faster with time
than the parton mean energy, which is expected since $\epsilon \equiv
n \left < E \right >$. 
We also see that the mean $\pt$ and mean transverse energy are
essentially the same at early times, and then they separate at late
times when parton mass (especially the strange quark mass) plays a
bigger role as the mean $\pt$ becomes smaller. 
Furthermore, the amount by which the parton mean energy 
is above the mean transverse energy reflects the contribution 
from the squared longitudinal momentum $\pz^2$ 
of partons under consideration (i.e. within $|\eta|<1/2$).
In addition, the two dotted lines in the left panel of
Fig.~\ref{fig:center} show the parton energy density and number
density when all partons (including unformed partons) are included. We
see that unformed partons contribute little to the energy density and number
density except for the earliest stage ($t<0.4$ fm/$c$ for partons
at mid-spacetime-rapidity in this study),  
and their relative contribution to the energy density is less than
that to the number density since the average formation time of hard
partons is shorter than that of soft partons in the string melting
version of the AMPT model \cite{Lin:2004en}. 

\subsection{Time evolution of effective temperatures}

The right panel of Fig.~\ref{fig:center} shows 
time evolutions of effective temperatures extracted from different
variables via Eqs.~(\ref{tpB}-\ref{teB}) for the center cell 
in central Au+Au collisions at $200A$ GeV. 
Here we see that the effective temperatures extracted from  
the parton energy density (filled circles), number density (solid 
curve), and transverse energy density (dashed curve) are all
significantly higher than those extracted from the shown mean momentum
variables, i.e., those extracted from the parton mean $\pt$ (open 
circles), mean $\pt^2$ (dotted curve), and mean 3-momentum (dot-dashed
curve). 
If the local parton system in this cell were in full chemical and
thermal equilibrium, temperature values extracted from these different
variables at a given time would all be the same and we would only see
one curve. Thus the different effective temperatures reflect the
non-equilibrium nature of the local parton system.
We also see that effective temperature $\te$ extracted from  
the parton energy density is very close to $\tet$ that is extracted
from the parton transverse energy density throughout the time evolution. 
While the effective temperature $\tmpt$ extracted from the parton mean
$\pt$ is very close to $\tmptsq$ that is extracted from the parton
mean $\pt^2$, it is higher than $\tmp$ (extracted from the parton mean 
3-momentum) at early times and then lower than $\tmp$ at late
times. This suggests that the mean squared longitudinal momentum 
$\pz^2$ is lower than the mean squared momentum along a transverse 
direction at early times but higher than that at later times; 
this is consistent with an earlier study on the pressure anisotropy in
the AMPT model \cite{Zhang:2008zzk}, where the ratio of the 
longitudinal pressure over the transverse pressure starts initially
below one but then quickly exceeds one after a time of several fm/$c$.
Recently pressure anisotropy has also been considered within the 
hydrodynamic framework 
\cite{Florkowski:2010cf,Martinez:2010sc,Bazow:2013ifa}.

From the right panel of Fig.~\ref{fig:center}, we also observe that 
$\te$ is mostly between $\tmpt$ and $\tn$, the effective temperature
extracted from the parton number density, 
while being close to $\tn$. 
This can be understood in the following. 
Relation $\epsilon_{\rm T}= n \left < E_{\rm T} \right >$ gives 
\ber
\tet^4 =\tn^3 ~ \tmet. 
\eer
Since $T_{\epsilon_{\rm T}} \simeq T_{\epsilon}$ as shown in the right
panel of Fig.~\ref{fig:center} and at high densities 
$\left < E_{\rm T} \right > \simeq \left < \pt \right >$ 
as shown in the left panel of Fig.~\ref{fig:center}, I can write
\ber
T_{\epsilon}^4 \simeq T_n^3 ~ T_{\left < \pt \right >}.
\label{trelation}
\eer
I find that the above relation for $\te$ is a good approximation 
at high densities: 
for all cells within $|\eta|<1/2$ in all four collisions systems that
I considered, the actual $\te$ value is found to be always within 6\%
of the value as given by Eq.~(\ref{trelation}) if $\te > 100$~MeV in
the cell.

\begin{figure}[h]
\includegraphics[width=6 in]{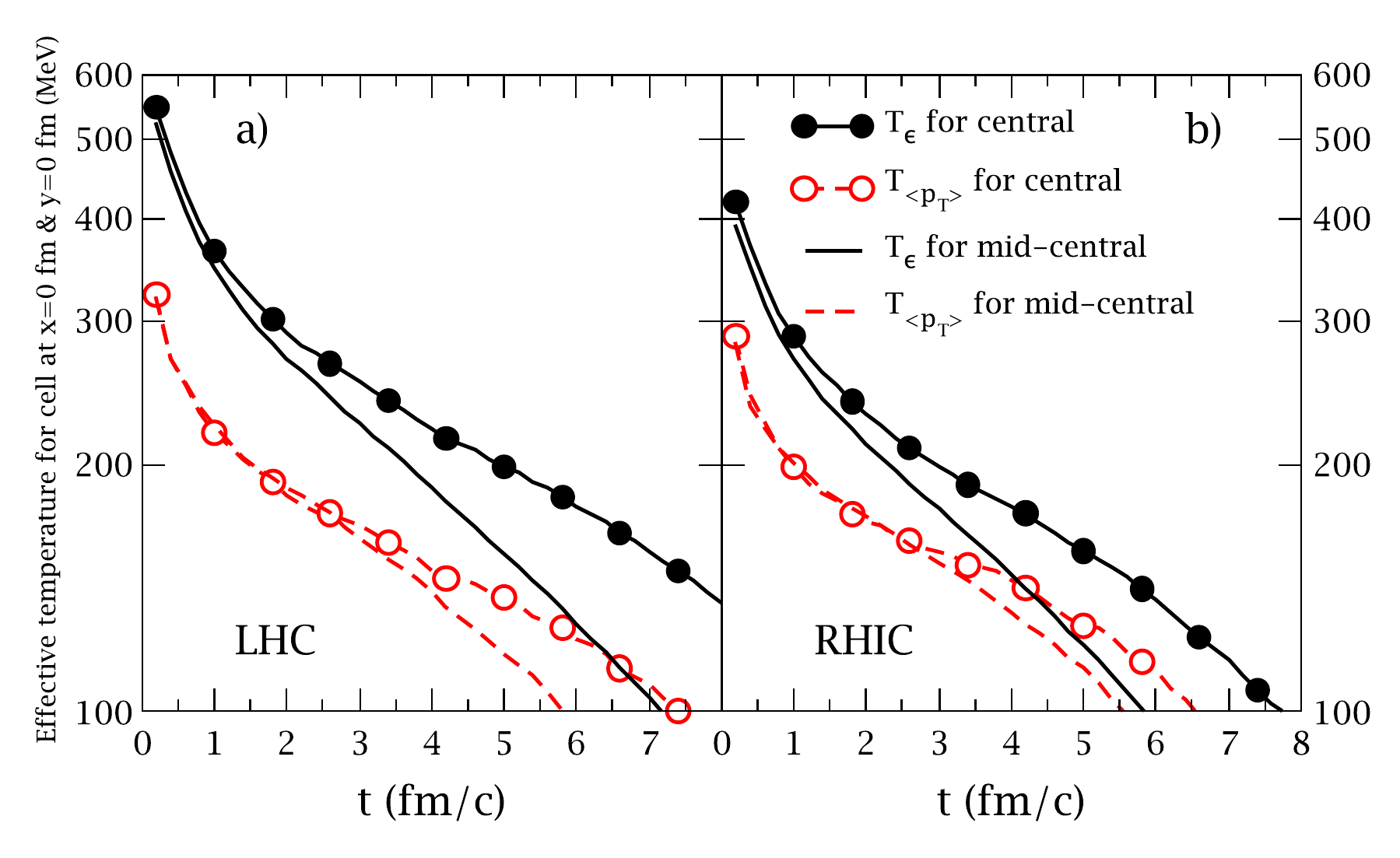}
\caption{(Color online) 
AMPT-SM results on effective temperatures as functions of time
for the center cell in 
a) Pb+Pb collisions at $2760A$ GeV,
and 
b) Au+Au collisions at $200A$ GeV. 
}
\label{fig:3t}
\end{figure}

Fig.~\ref{fig:3t} shows the time evolutions of $\te$ and $\tmpt$ in
the center cell in these different collisions, and we see $\te>\tmpt$
in each case.  
It is also seen that the initial temperature $\te$ (and $\tmpt$) in
mid-central collisions is similar to that in central collisions at
the same energy, but the temperature in mid-central collisions
decreases faster with time. 
On the other hand, the initial temperature $\te$ (and $\tmpt$) in LHC
collisions is higher than that in RHIC collisions at the same
centrality, but the shape of the temperature evolution with time
looks similar at these two energies.

\begin{figure}[h]
\includegraphics[width=4 in]{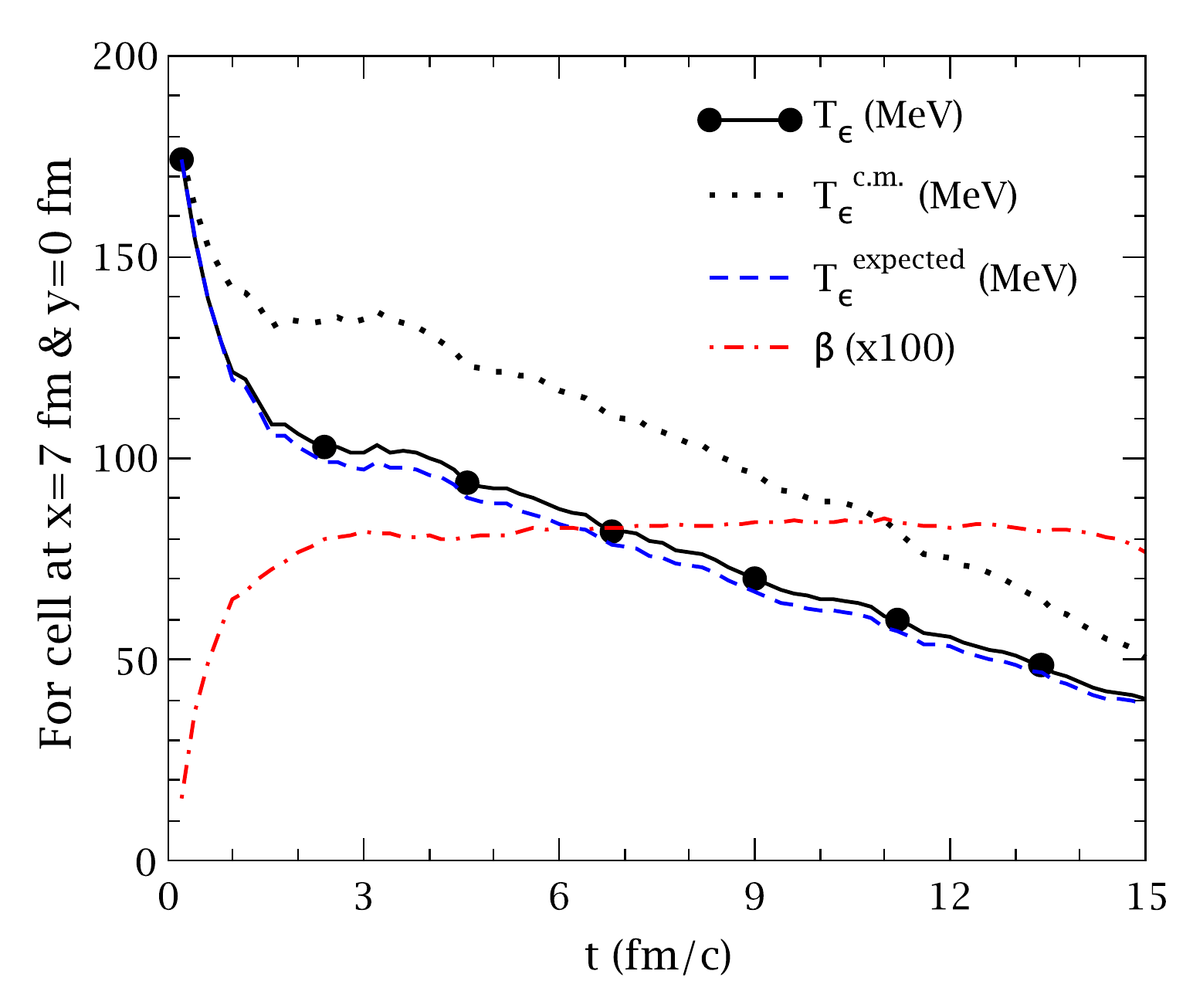}
\caption{(Color online) 
AMPT-SM results on effective temperatures as functions of time 
for the cell at $x=7$~fm \& $y=0$~fm in 
central Au+Au collisions at $200A$ GeV; the dot-dashed curve
represents the flow magnitude (multiplied by 100) in the cell. 
}
\label{fig:checkt}
\end{figure}

I have also checked the relation between $\te$ extracted from the 
parton energy density in the rest frame of the cell and $\tecm$, 
which is the ``apparent'' temperature if one simply converts the 
energy density of the cell in the center-of-mass (c.m.) frame into
temperature using Eq.~(\ref{teB}). 
Fig.~\ref{fig:checkt} shows 
the time evolutions of $\te$ (filled circles)  
and $\tecm$ (dotted curve) in the cell at $x=7$~fm \& $y=0$~fm in 
central Au+Au collisions at $200A$ GeV,
where the ``apparent'' temperature $\tecm$ is much higher than
$\te$ after 1 fm/$c$. 
For a volume cell that can be described by energy density $\epsilon$
and pressure $P$, the energy density $\epsilon$ in its rest frame is
related to the energy density $\epcm$ in the c.m. frame by 
\ber
\epcm=\gamma^2 (\epsilon+P)-P,
\eer
where $\gamma$ is the Lorentz factor for the moving cell in the c.m.
frame. Using $P=\epsilon/3$ for a massless parton gas, I get
\ber
\epsilon&=&\epcm \left (\frac {3}{4\gamma^2-1} \right ), \nonumber \\
\te^{\rm expected}&=&\tecm \left ( \frac {1-\beta^2} {1+\beta^2/3} \right )^{1/4}, 
\label{tecm}
\eer
where $\beta$ is the magnitude of the flow of the cell in the c.m.
frame. Using the flow magnitude $\beta$ shown as the dot-dashed curve
in Fig.~\ref{fig:checkt}, Eq.~(\ref{tecm}) gives $\te^{\rm expected}$
(the expected value of $\te$) as the dashed curve,  
which is very close to the actual $\te$ calculated directly from the
energy density in the rest frame of the cell.
The small difference between $\te$ and $\te^{\rm expected}$ 
may come from the fact that partons are not massless and that
pressures along different axes are not fully isotropicalized 
during the evolution \cite{Zhang:2008zzk,Xu:2007aa}.

\subsection{Spatial dependence of effective temperatures}

\begin{figure}[h]
\includegraphics[width=6 in]{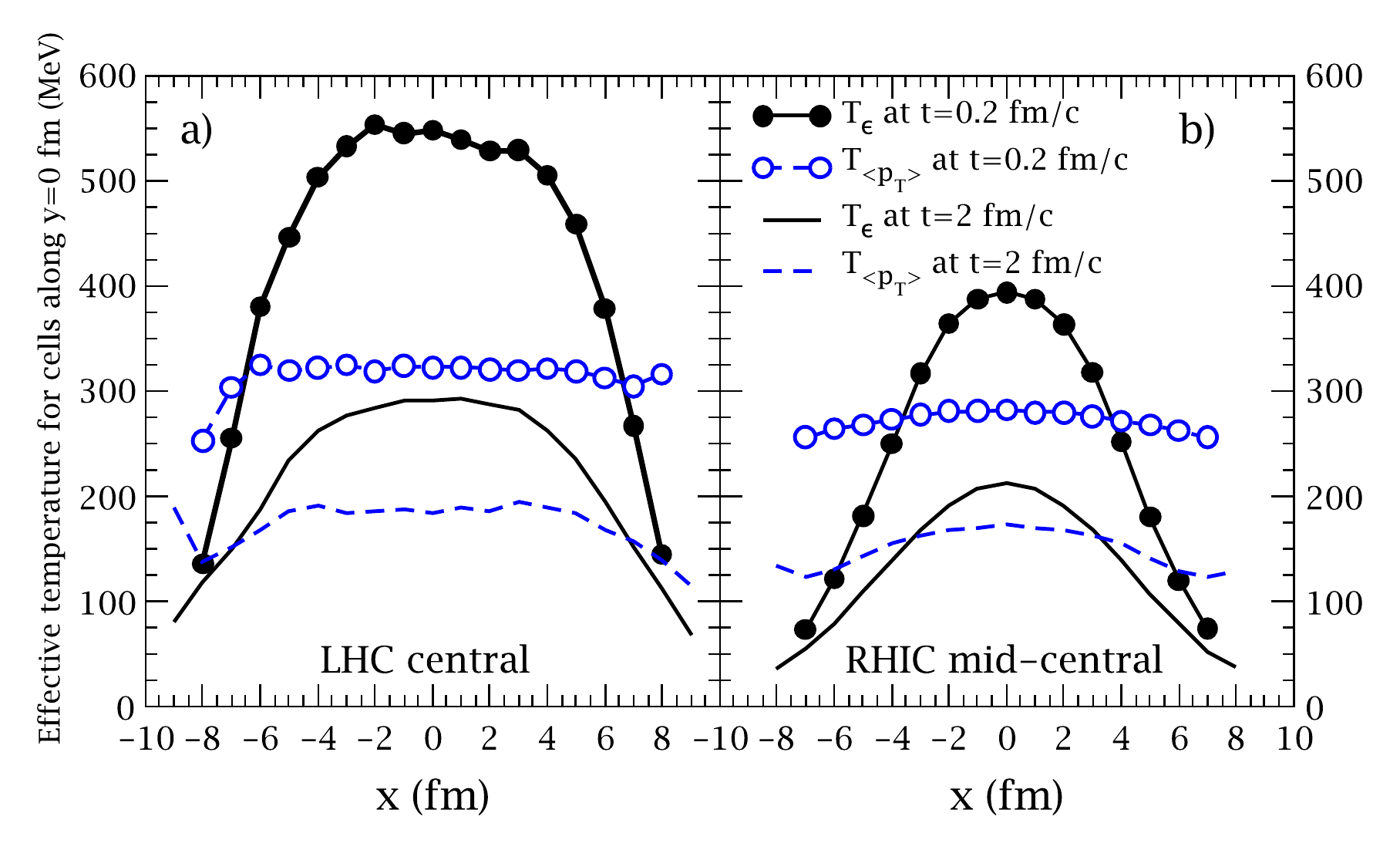} 
\caption{(Color online) 
AMPT-SM results on effective temperatures in cells along the impact
parameter at $t=0.2$ and $2.0$ fm/$c$: 
a) for central Pb+Pb collisions at $2760A$ GeV,
and b) for mid-central Au+Au collisions at $200A$ GeV.
}
\label{fig:compare1}
\end{figure}

\begin{figure}[h]
\includegraphics[width=6 in]{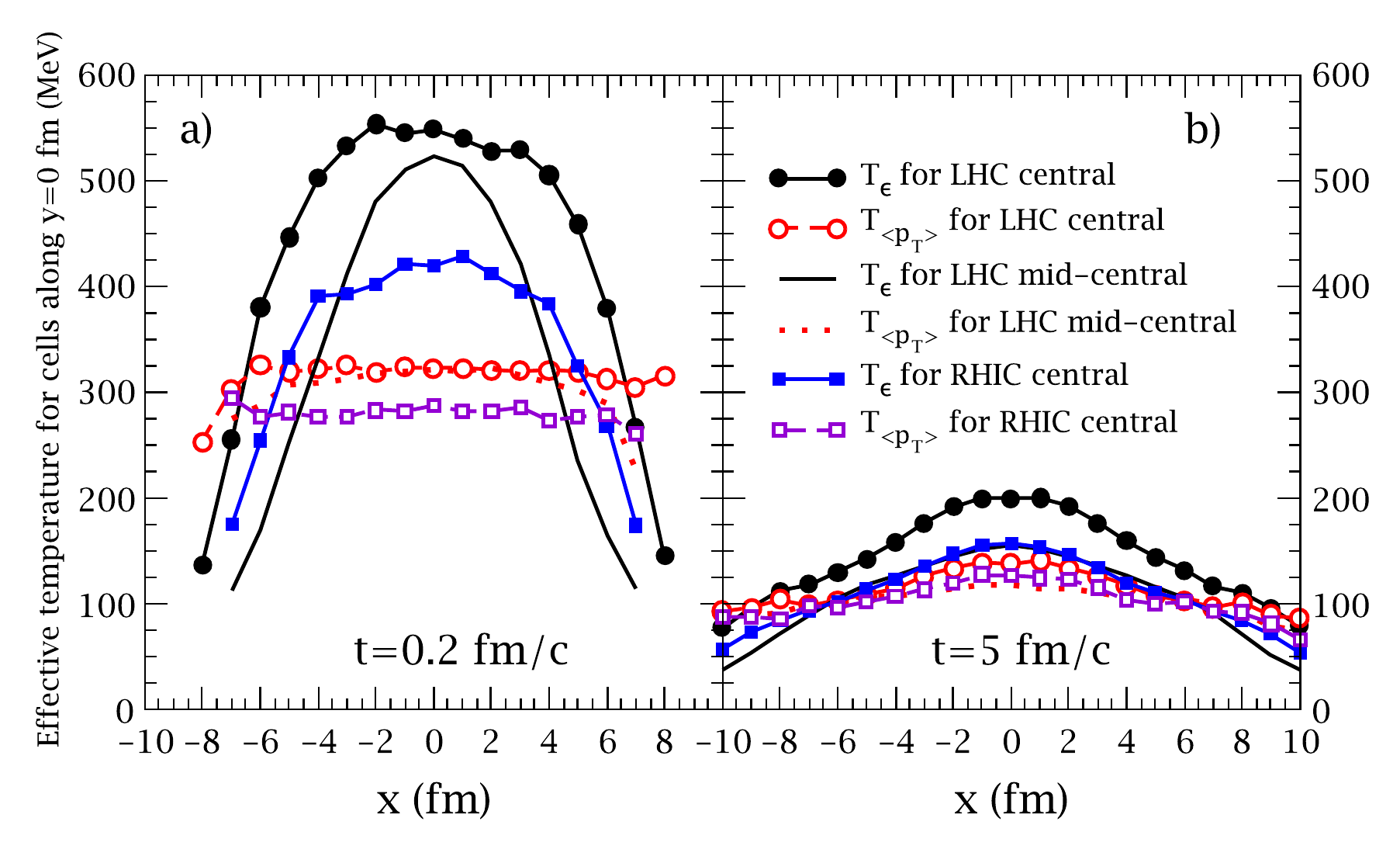} 
\caption{(Color online) 
AMPT-SM results on effective temperatures in cells along the impact
parameter for different collisions: 
a) at $t=0.2$ fm/$c$, and b) at $t=5.0$ fm/$c$.
}
\label{fig:compare2}
\end{figure}

I now look at how the effective temperatures depend on location in
the overlap volume. 
Fig.~\ref{fig:compare1} shows how $\te$ and $\tmpt$ change with
location along the impact parameter axis at different times:
$t=0.2$ and $2.0$ fm/$c$. 
In both central Pb+Pb collisions at $2760A$ GeV  
and mid-central Au+Au collisions at $200A$ GeV,  
we see that the initial effective temperatures $\tmpt$ at $t=0.2$
fm/$c$ change little over most of the overlap volume. 
We also see that $\te > \tmpt$ over the inner part of the overlap
volume, e.g., within $|x|<6$ fm for central Pb+Pb collisions at
$2760A$ GeV and within $|x|<3$ fm for mid-central Au+Au collisions at
$200A$ GeV at $t=0.2$ fm/$c$. 
Note that, in order to limit the statistical fluctuations,
volume cells that have less than 20 partons after summing
over simulated events are not shown in
Figs.~\ref{fig:compare1}-\ref{fig:compare2}. 

The spatial dependences of effective temperatures $\te$ and $\tmpt$ in
three different collision systems are shown in Fig.~\ref{fig:compare2}a
for $t=0.2$ fm/$c$ and in Fig.~\ref{fig:compare2}b for $t=5.0$
fm/$c$. From Fig.~\ref{fig:compare2}a 
I find that the initial effective temperatures $\tmpt$ 
for central and mid-central Pb+Pb collisions at $2760A$ GeV 
are almost the same in the overlap volume except that
the spatial width is smaller for mid-central collisions.
In contrast, the initial effective temperatures $\te$ for central LHC
collisions are higher and much wider in spatial width 
than those in mid-central LHC collisions. 
These features are also seen in results at the RHIC energy. 
Furthermore, although the initial temperatures $\te$ in 
mid-central Pb+Pb collisions at $2760A$ GeV are higher than those in
central Au+Au collisions at $200A$ GeV in the inner part of the overlap
volume, the width of $\te$ along the impact parameter axis is
smaller. From Fig.~\ref{fig:compare2}b one sees that the effective
temperature $\te$ at $t=5.0$ fm/$c$ for central LHC collisions 
are overall significantly higher than the other collisions systems. 
In addition, temperatures $\te$ (and $\tmpt$) at $t=5.0$ fm/$c$ 
are mostly similar for mid-central Pb+Pb collisions at $2760A$ GeV and
central Au+Au collisions at $200A$ GeV, suggesting that the lifetime of
the quark-gluon plasma is similar in these two collision systems.

\subsection{Over-population of partons}

We have already seen from Figs.~\ref{fig:compare1}-\ref{fig:compare2} 
that $\te > \tmpt$ over the inner part of the overlap volume in
these high energy heavy ion collisions, and this is essentially due to 
$\tn > \tmpt$ according to Eq.~(\ref{trelation}). 
This reflects the fact that the parton system in a subvolume of the
overlap region in such a collision, even after being averaged over
many events at the same impact parameter, is not in full
chemical equilibrium as defined for an ideal quark-gluon plasma. 
The relation $\tn > \tmpt$ means that for 
the parton number density we have $n > n(\tmpt)$, where $n(\tmpt)$ 
is the parton number density expected for a quark-gluon plasma  
in full chemical equilibrium at temperature $\tmpt$. For example, $\te
\sim 1.7~\tmpt$ in Figs.~\ref{fig:compare1}a for $x=0$ fm at $t=0.2$
fm/$c$ in central LHC collisions means that, according to
Eq.~(\ref{teB}) and the approximation of Eq.~(\ref{trelation}), $n
\sim 1.7^4~n(\tmpt) \simeq 8.4~n(\tmpt)$ for that 
center cell at that time. 
Later at $t=5.0$ fm/$c$, we see in Figs.~\ref{fig:compare1}b that
$\te \sim 1.44~\tmpt$, which gives $n \sim 4.3~n(\tmpt)$ at that
time. Therefore one may say that the parton system in that subvolume
is over-populated, in that the parton density is too high compared to
that expected for an ideal QGP that has the same parton mean
transverse momentum.

\begin{figure}[h]
\includegraphics[width=6 in]{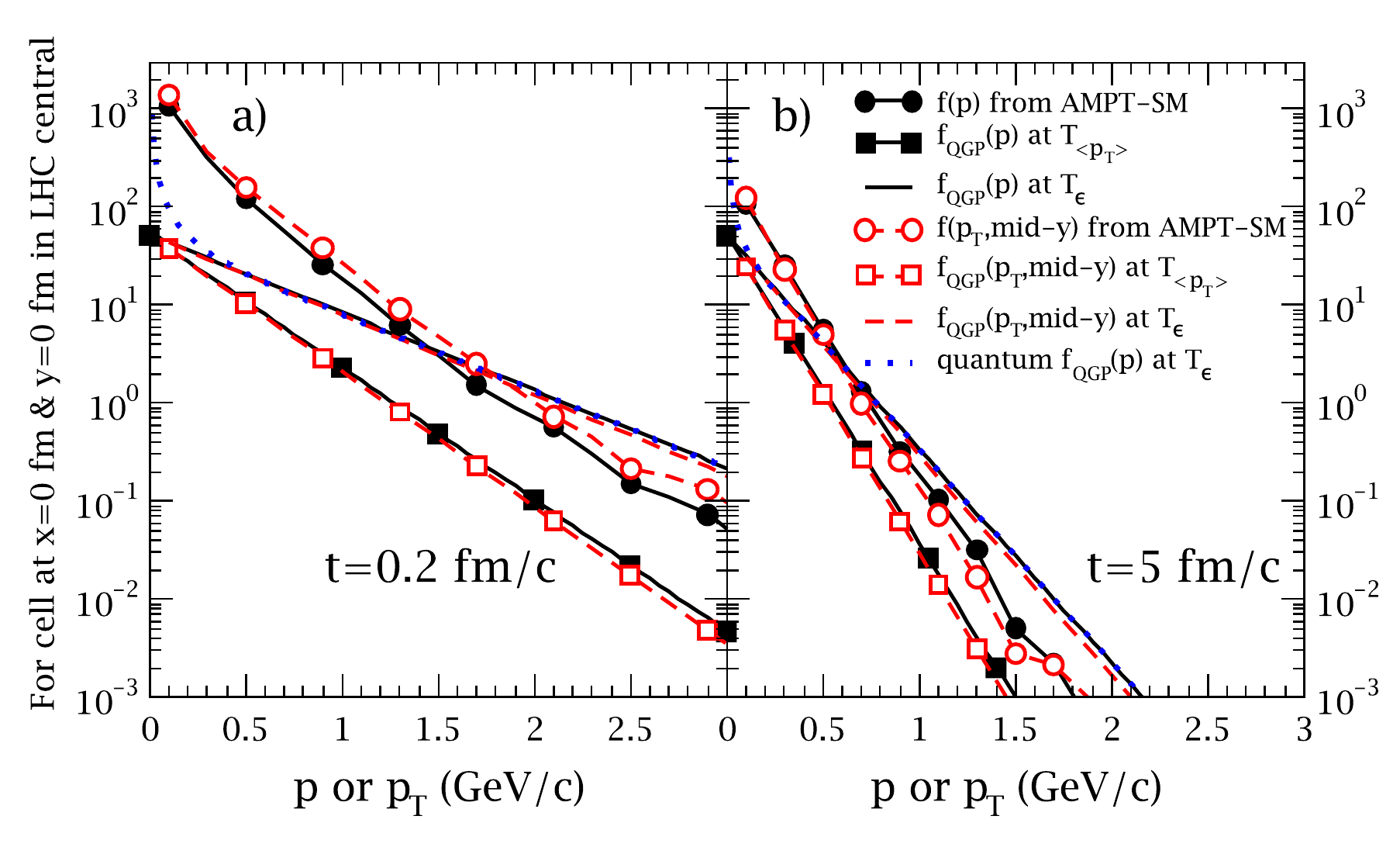}
\caption{(Color online)  
Phase-space distribution of partons in the center cell in central
Pb+Pb collisions at $2760A$ GeV versus momentum $p$ (solid curves) 
or versus $\pt$ at mid-rapidity (dashed curves) 
a) at $t=0.2$ fm/$c$, and b) at $t=5.0$ fm/$c$. 
Circles represent results from the AMPT-SM model, 
curves with squares represent an ideal QGP at temperature $\tmpt$, 
solid and dashed curves without symbols represent an ideal QGP at
temperature $\te$, and dotted curves represent an ideal QGP at
temperature $\te$ with quantum statistics.
}
\label{fig:overp1}
\end{figure}

To further illustrate this, I calculate the phase-space density of
partons. In terms of the magnitude of parton 3-momentum in the
rest frame of a cell, I have
\ber 
f(p) \equiv (2 \pi)^3 \frac {dN}{d^3x d^3p} =2 \pi^2 \frac{dN/d^3x}{p^2 dp}.
\eer
I also calculate the phase-space density in terms of parton $\pt$ at
a given rapidity $y$ in the rest frame of a cell, where one can write 
\ber  
f(\pt,y) =4 \pi^2 \frac{dN/d^3x}{\pt d\pt d \pz}. 
\label{fpt1}
\eer 
I choose mid-rapidity partons, i.e., partons within $|y|<1/2$ in the
rest frame of the cell. 
By approximating the parton system as massless, which
is a reasonable approximation when the temperature is high 
(see left panel of Fig.~\ref{fig:center}), I obtain
\ber 
f(\pt, {\rm mid\!-\!y}) 
\simeq 4 \pi^2 \frac{dN/d^3x}{\pt d\pt \Delta \pz}
\simeq \frac {4 \pi^2}{2 \sinh(1/2)} \frac{dN/d^3x}{\pt^2 d\pt}, 
\eer
where $\Delta \pz$ represents the longitudinal momentum range under
consideration. 
For comparison, the expected phase-space distribution for an ideal
QGP at temperature $T$ is given by 
\ber 
\fqgp(p) = \gb e^{-E/T} \simeq \gb e^{-p/T},
\fqgp(\pt,y)  = \gb e^{-m_{\rm T} \cosh(y)/T},
\eer
when the Boltzmann distribution is used for partons that are assumed
to be massless. I then make the following approximation for
mid-rapidity:
\ber
\fqgp(\pt, {\rm mid\!-\!y}) \simeq \gb e^{-\pt \cosh(1/4)/T}.
\eer

The AMPT-SM results for the center cell at two different times in
central Pb+Pb collisions at $2760A$ GeV are shown in Fig.~\ref{fig:overp1}, 
in comparison with the expected phase-space distributions for an ideal
QGP at the corresponding temperature $\te$ (548~MeV at $t=0.2$ fm/$c$
and 199~MeV at $t=5$ fm/$c$) or $\tmpt$ (322~MeV at $t=0.2$ fm/$c$ and
138~MeV at $t=5$ fm/$c$). 
We first see that the AMPT-SM results (curves with circles) and the
expected curves for an ideal QGP at temperature $\tmpt$ (curves with
squares) at a given time have a similar overall slope. 
This reflects the fact that the effective temperature $\tmpt$ is
only determined by the parton mean $\pt$ in the cell; it also 
indicates that the $\pt$ distribution of partons in the subvolume
here is not too far from a thermal distribution. 
In the meantime, the AMPT-SM curves are much higher in magnitude
than the ideal QGP curves at temperature $\tmpt$, suggesting that the
parton matter is over-populated by a large factor.
When compared with the expected distributions for an ideal QGP at
temperature $\te$ (solid and dashed curves without symbols), 
partons in the AMPT-SM results are still over-populated below a given
value of $\pt$ or $p$. 
This value is smaller at a later time as shown in
Fig.~\ref{fig:overp1}: partons in that center cell are seen to be
over-populated below $\pt \simeq 1.4$ GeV/$c$ at $t=0.2$ fm/$c$ but
below $\pt \simeq 0.6$ GeV/$c$ at $t=5$ fm/$c$, when compared with 
an ideal QGP at temperature $\te$.

From Fig.~\ref{fig:overp1} we see that the phase-space distributions
in momentum $p$ or $\pt$ are quite close to each other. At
zero-momentum, 
both $\fqgp(p)$ and $\fqgp(\pt,{\rm mid\!-\!y})$ approach the value
$\gb=52$ (the degeneracy factor of QGP for $N_f=3$) 
for an ideal QGP with Boltzmann statistics. 
Considering Fermi-Dirac statistics,  the phase-space densities of
quarks and anti-quarks  are bound to be below one for each degree of
freedom due to the Pauli exclusion principle, therefore the phase
space density contributed from quarks and anti-quarks must be below
$12N_f=36$ (their total degeneracy  factor for $N_f=3$) regardless of
temperature.  
Therefore the high phase-space densities above this value shown in 
Fig.~\ref{fig:overp1} cannot be explained by a full population of the
phase space by quarks and antiquarks.
It cannot be explained by the Bose-Einstein distribution of gluons in
an ideal QGP either. 
Dotted curves in Fig.~\ref{fig:overp1} represent the expected quantum
phase-space distributions at temperature $\te$ 
when Bose-Einstein and Fermi-Dirac distributions are used for a 
massless quark-gluon plasma:
\ber 
\fqgp^{quantum}(p) =\frac {16} {e^{p/T}-1}+\frac {12N_f} {e^{p/T}+1}.
\label{fquantum}
\eer
For an ideal QGP, we see that the expected quantum phase-space
densities are very close to the expected Boltzmann densities except at
very low $\pt$, and the AMPT-SM results are still much higher than the
quantum phase-space densities except at extremely low $\pt$. 
Since there is no bound for the gluon phase-space densities due to
quantum statistics, one may argue that the high density regions as
shown in Fig.~\ref{fig:overp1} indicate that at least gluons are
over-populated there. This may be analogous to the glasma produced
from the color-glass-condensate \cite{Blaizot:2011xf,Lappi:2006fp}. 
 
Therefore, when the parton system in a volume cell satisfies $n >
n(\tmpt)$, I consider partons in that cell as being over-populated 
relative to an ideal QGP at temperature $\tmpt$. 
So I use the AMPT-SM results to investigate 
regions within $|\eta|<1/2$ that have over-populated partons. 
I first define a ``critical'' temperature $T_c$ above which a 
dense matter can be considered as being well inside the QGP phase,
then a cell that satisfies the condition $\te > T_c$ is called a QGP
cell, and a QGP cell that also satisfies the condition $\te > \tmpt$
is called an over-populated cell. In this study I take $T_c=150$~MeV,
which corresponds to a ``critical'' energy density $\epsilon_c=1.05$
GeV/fm$^3$ for $N_f=3$ as given by Eq.~(\ref{teB}). 

\begin{figure}[h]
\includegraphics[width=4 in]{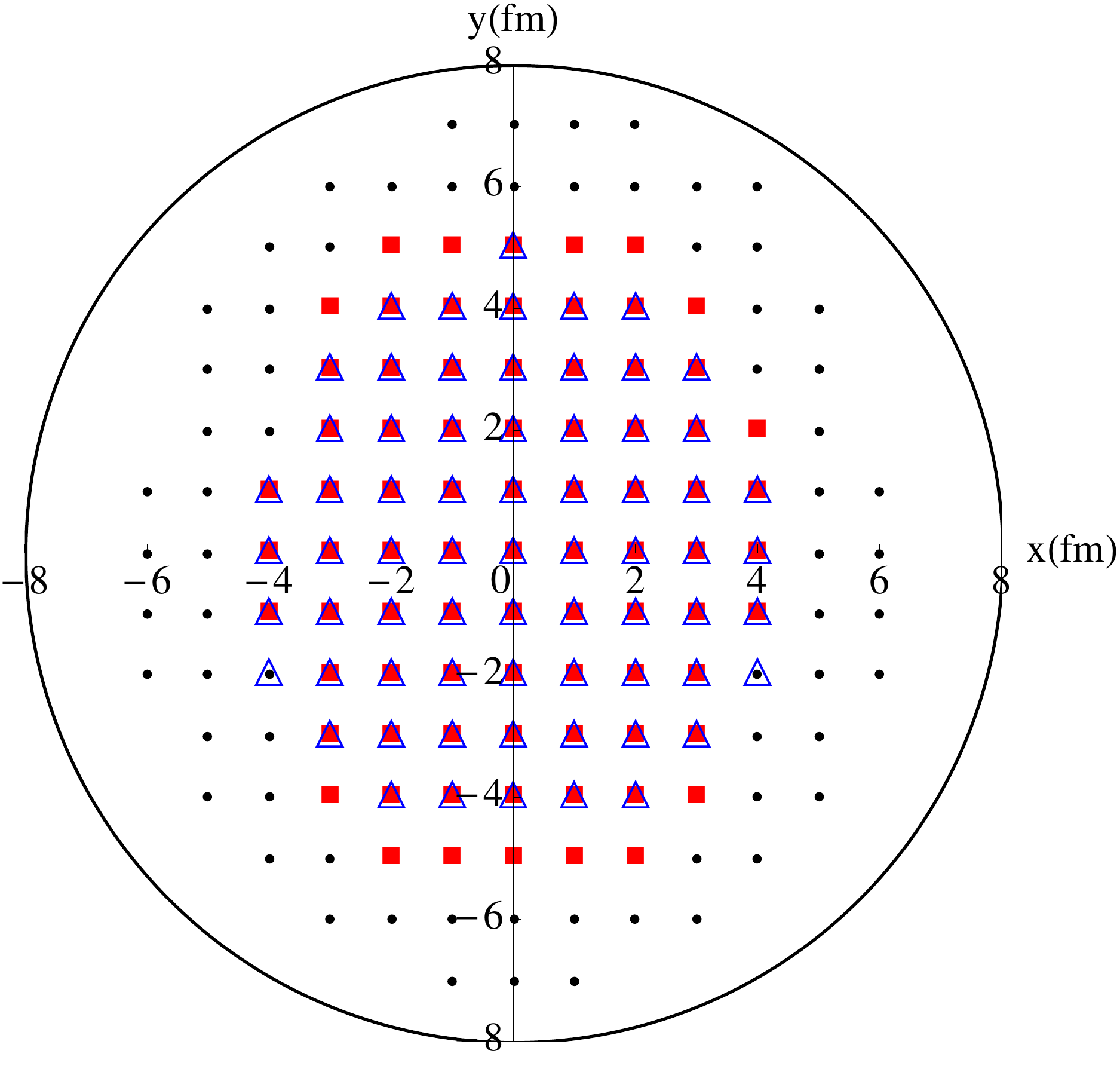}
\caption{(Color online)  
For partons within $|\eta|<1/2$ in mid-central Pb+Pb collisions at 
$2760A$ GeV,  
filled circles represent the center locations in the transverse plane  
for QGP cells at $t=0.2$ fm/$c$, filled squares represent
over-populated QGP cells at $t=0.2$ fm/$c$, and open triangles
represent over-populated QGP cells at $t=3.0$ fm/$c$. The circle is 
drawn as a reference shape.
}
\label{fig:overp2}
\end{figure}

Fig.~\ref{fig:overp2} shows the transverse locations of QGP cells 
(i.e. with $\te >T_c$) and over-populated cells (i.e. with $\te >T_c$
and $\te >\tmpt$)  in mid-central Pb+Pb collisions at $2760A$ GeV. 
We can see that at $t=0.2$ fm/$c$ the initial extension of QGP cells 
(filled circles) along the $y$-axis is bigger than that along the
$x$-axis due to the spatial asymmetry in mid-central collisions.
More that half of these QGP cells are over-populated at
this early time, as indicated by filled squares, and they are located
in the inner part of the overlap volume. 
Later at $t=3.0$ fm/$c$, QGP cells that are over-populated (open
triangles) still occupy an area in the transverse plane almost as
large as that at $t=0.2$ fm/$c$ (filled squares),   while the shape of
overpopulated cells in the transverse plane  at $t=3.0$ fm/$c$ has a
smaller spatial asymmetry than that at  $t=0.2$ fm/$c$.

\begin{figure}[h]
\includegraphics[width=4 in]{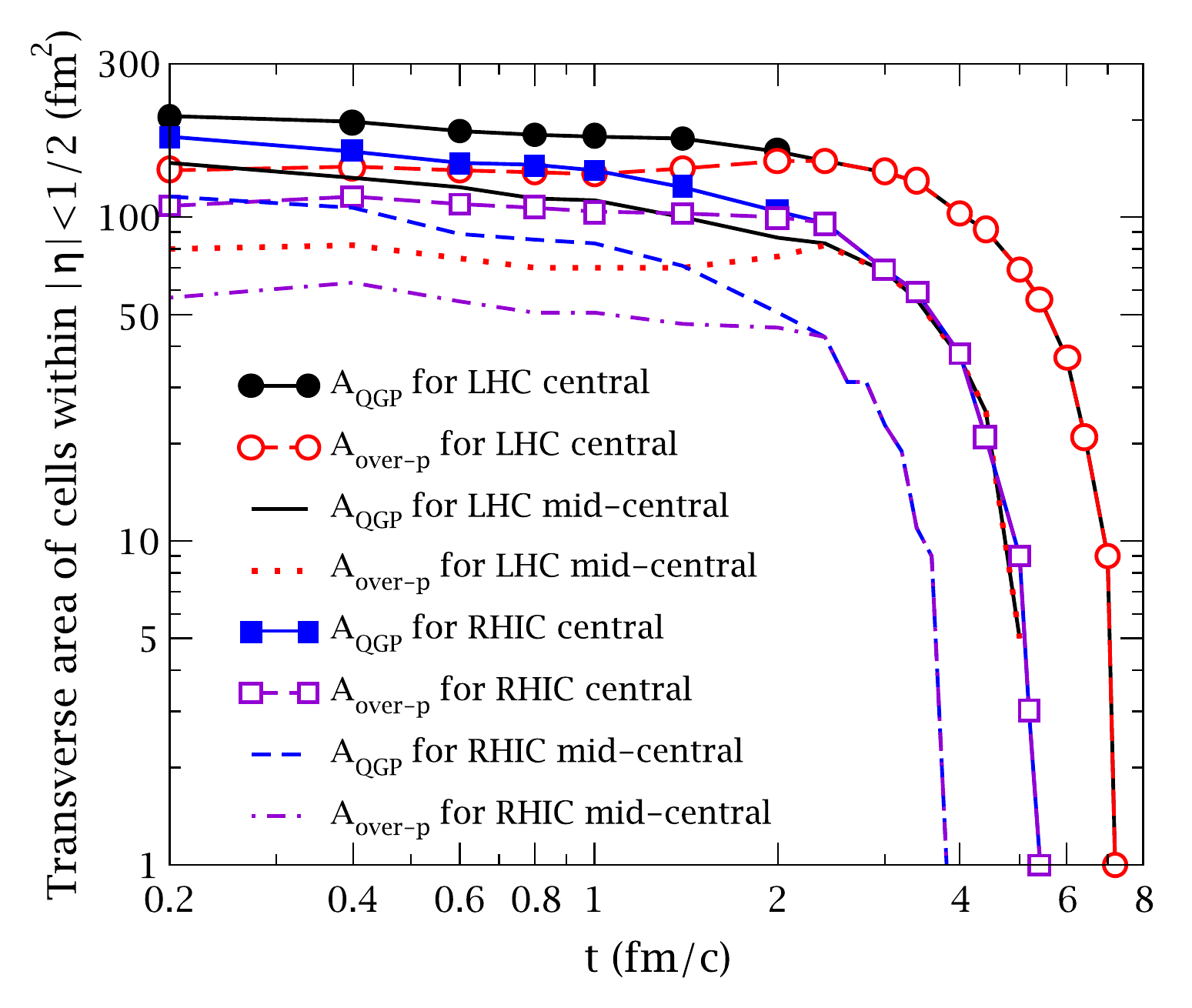}
\caption{(Color online) 
Transverse areas of QGP cells ($A_{\rm QGP}$) and transverse areas of
QGP cells that are over-populated ($A_{\rm over-p}$) as
functions of time for different collisions.
}
\label{fig:overp3}
\end{figure}

In Fig.~\ref{fig:overp3} I show the transverse areas of QGP cells
and the transverse areas of over-populated QGP cells 
as functions of time in different collisions. Note that each cell here
covers $|\eta|<1/2$ along the beam direction. 
For all four collision systems, we see that initially a significant
fraction, roughly from 50\% to 70\%, of the QGP cells are
over-populated with partons. This fraction is higher for central
collisions than mid-central collision at the same energy and also
higher for the LHC energy than the RHIC energy at the same
centrality. After 2-3 fm/$c$, however, all QGP cells are
over-populated. 
Although the transverse area of QGP cells decreases monotonously over
time, the transverse area of over-populated cells does not change much
and may even slightly increase over the first several fm/$c$ of
time. One also sees that the areas in central RHIC collisions are
mostly bigger than the corresponding areas in mid-central LHC
collisions even though the peak temperature in central RHIC
collisions is lower (as shown in  Fig.~\ref{fig:compare2}a), while
the QGP phases in the two collision systems show a similar lifetime.

\subsection{Describing parton
  over-population with parton phase-space occupancy factors} 

We can represent the over-population of partons with parton phase-space 
occupancy factors. Let us introduce the following quantum
phase-space distribution function for QGP that has a zero baryon
chemical potential but is off chemical equilibrium:
\ber 
\fqgp^{non-eq}(p) =\frac {16 \gamma_g} {e^{p/T}-1}+\frac {12N_f \gamma_q} {e^{p/T}+1},
\label{fnoneq}
\eer
where $\gamma_g$ is the gluon phase-space occupancy and 
$\gamma_q$ is the quark (and anti-quark) phase-space occupancy.
The above distribution gives the energy density as
$\epsilon^{non-eq}(T)=\pi^2(16 \gamma_g+10.5N_f \gamma_q) T^4/30$.
Since the 
effective temperature $\tmpt$ represents the shape of the phase-space
distribution, it will be used as temperature $T$ in
Eq.~(\ref{fnoneq}), and matching $\epsilon^{non-eq}(\tmpt)$ to the
local energy density value $\epsilon \equiv 3\gb \te^4/\pi^2$ enables
us to extract the gluon phase-space occupancy of each volume cell as
\ber 
\gamma_g=\frac {3}{32} \left [ \frac{240 (4+3N_f)}{\pi^4} \left ( \frac{\te}{\tmpt} \right )^4-7N_f \gamma_q \right ].
\label{mug}
\eer
Alternatively we can use the following Boltzmann distribution function
for a massless QGP that is off chemical equilibrium:
\ber
\fqgpb^{non-eq}(p)=16 \gamma_g e^{-p/T}+12N_f \gamma_q e^{-p/T}, 
\label{fnoneqB}
\eer
which leads to the energy density as
$\epsilon_B^{non-eq}(T)=12(4\gamma_g+3N_f \gamma_q)T^4/\pi^2$.
Matching $\epsilon_B^{non-eq}(\tmpt)$ to the local energy density value 
$\epsilon \equiv 3\gb \te^4/\pi^2$ then yields the gluon phase-space
occupancy for Boltzmann distributions as
\ber 
\gamma_g=\frac {1}{4} \left [ (4+3N_f) \left ( \frac{\te}{\tmpt} \right )^4-3N_f \gamma_q \right ].
\label{mugB}
\eer

As the gluon and quark composition of the partonic matter cannot be
addressed currently via the AMPT-SM model, 
the relationship between $\gamma_g$ and $\gamma_q$ is unknown. 
However, since $\gamma_q \in [0,1]$ due to the Pauli exclusion principle, we
can obtain the range of $\gamma_g$ as
$[\gamma_g^{min},\gamma_g^{max}]$, where 
$\gamma_g^{min}$ is the value of $\gamma_g$ for $\gamma_q=1$ and 
$\gamma_g^{max}$ is the value of $\gamma_g$ for $\gamma_q=0$.

\begin{figure}[h]
\includegraphics[width=6 in]{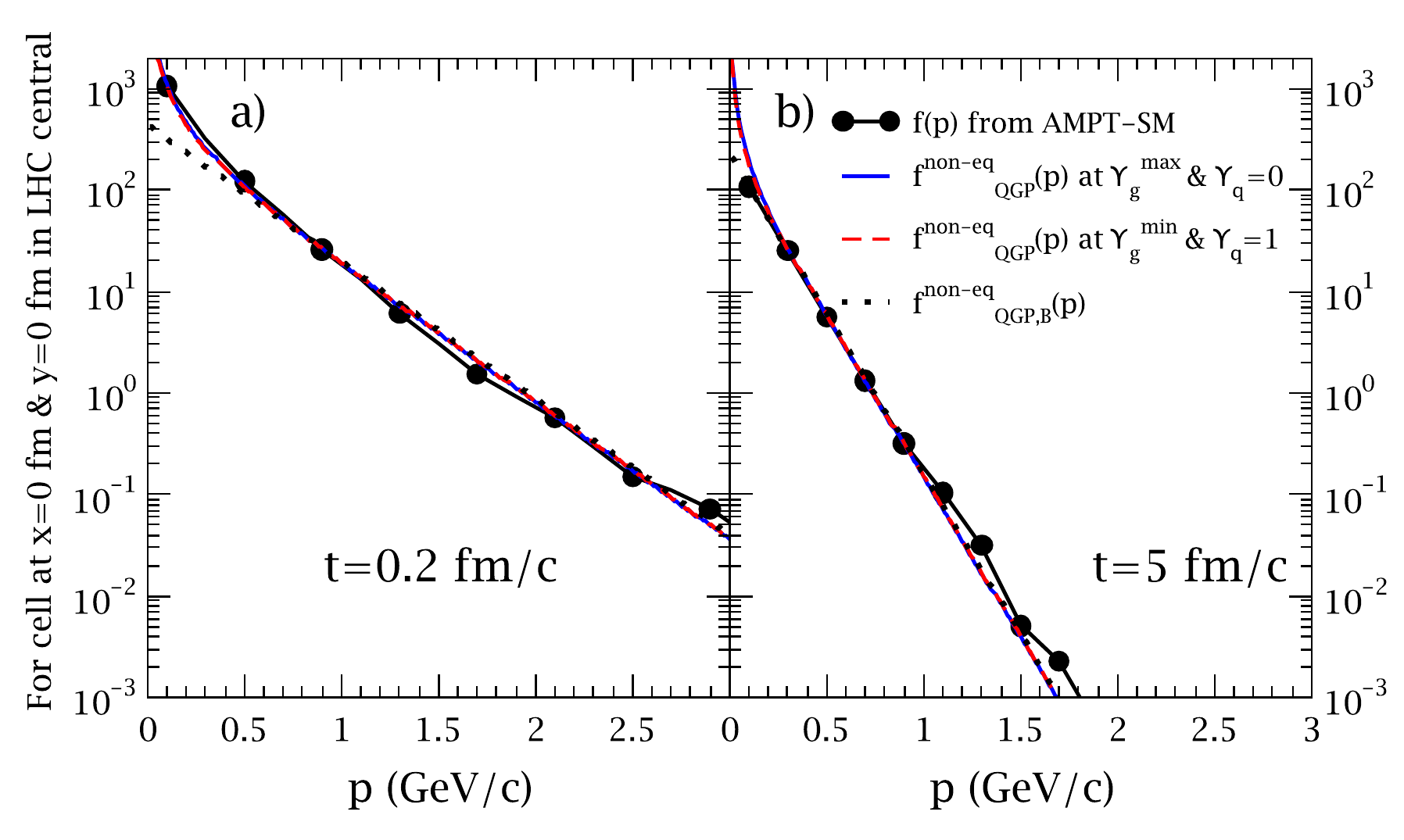}
\caption{(Color online)  
Phase-space distribution of partons in the center cell in central
Pb+Pb collisions at $2760A$ GeV versus momentum $p$
from AMPT-SM 
a) at $t=0.2$ fm/$c$, and b) at $t=5.0$ fm/$c$,  
in comparison with the corresponding non-equilibrium quantum 
distributions of Eq.~(\ref{fnoneq}) at $\gamma_g=\gamma_g^{max}$ \& $\gamma_q=0$ 
and at $\gamma_g=\gamma_g^{min}$ \& $\gamma_q=1$ 
and the non-equilibrium Boltzmann  
distribution of Eq.~(\ref{fnoneqB}).
}
\label{fig:fmug}
\end{figure}

Fig.~\ref{fig:fmug} shows the phase-space distributions of partons 
in the center cell of central Pb+Pb collisions at $2760A$ GeV 
from the AMPT-SM results at two different times (curves with circles).  
Also shown for comparisons are the quantum phase-space distributions of
Eq.~(\ref{fnoneq}) (solid curves and dashed curves) and the Boltzmann
distribution of Eq.~(\ref{fnoneqB}) (dotted curves) for QGP in
non-chemical-equilibrium at the corresponding temperature
$\tmpt$ (322~MeV at $t=0.2$ fm/$c$ and 138~MeV at $t=5$ fm/$c$). 
We see that all three non-chemical-equilibrium distributions in
Fig.~\ref{fig:fmug} describe the AMPT-SM results much better than the
chemical-equilibrium phase-space distributions shown in Fig.~\ref{fig:overp1}.  
We also see that the non-equilibrium quantum phase-space distributions 
for two sets of parton phase-space occupancies,
($\gamma_g=\gamma_g^{max},\gamma_q=0$) and ($\gamma_g=\gamma_g^{min},
\gamma_q=1$), are essentially the same; while the range of $\gamma_g$
for the non-equilibrium quantum distribution is $[23.1, 25.1]$ for
Fig.~\ref{fig:fmug}a and $[11.1, 13.1]$ for Fig.~\ref{fig:fmug}b.  
Since the Boltzmann distribution does not distinguish between bosons
and fermions, the non-equilibrium Boltzmann phase-space distribution 
$\fqgpb^{non-eq}(p)$ (dotted curves) is exactly the same for phase-space
occupancies ($\gamma_g=\gamma_g^{max},\gamma_q=0$) and
($\gamma_g=\gamma_g^{min}, \gamma_q=1$); 
while the range of $\gamma_g$ for the non-equilibrium Boltzmann distribution
is $[24.9, 27.2]$ for Fig.~\ref{fig:fmug}a and $[11.9, 14.2]$ for
Fig.~\ref{fig:fmug}b. 
We see that the gluon phase-space occupancy can be much bigger than
one. In addition, using the non-equilibrium quantum or Boltzmann
distribution does not affect the extracted range of $\gamma_g$ too much.

\begin{figure}[h]
\includegraphics[width=6 in]{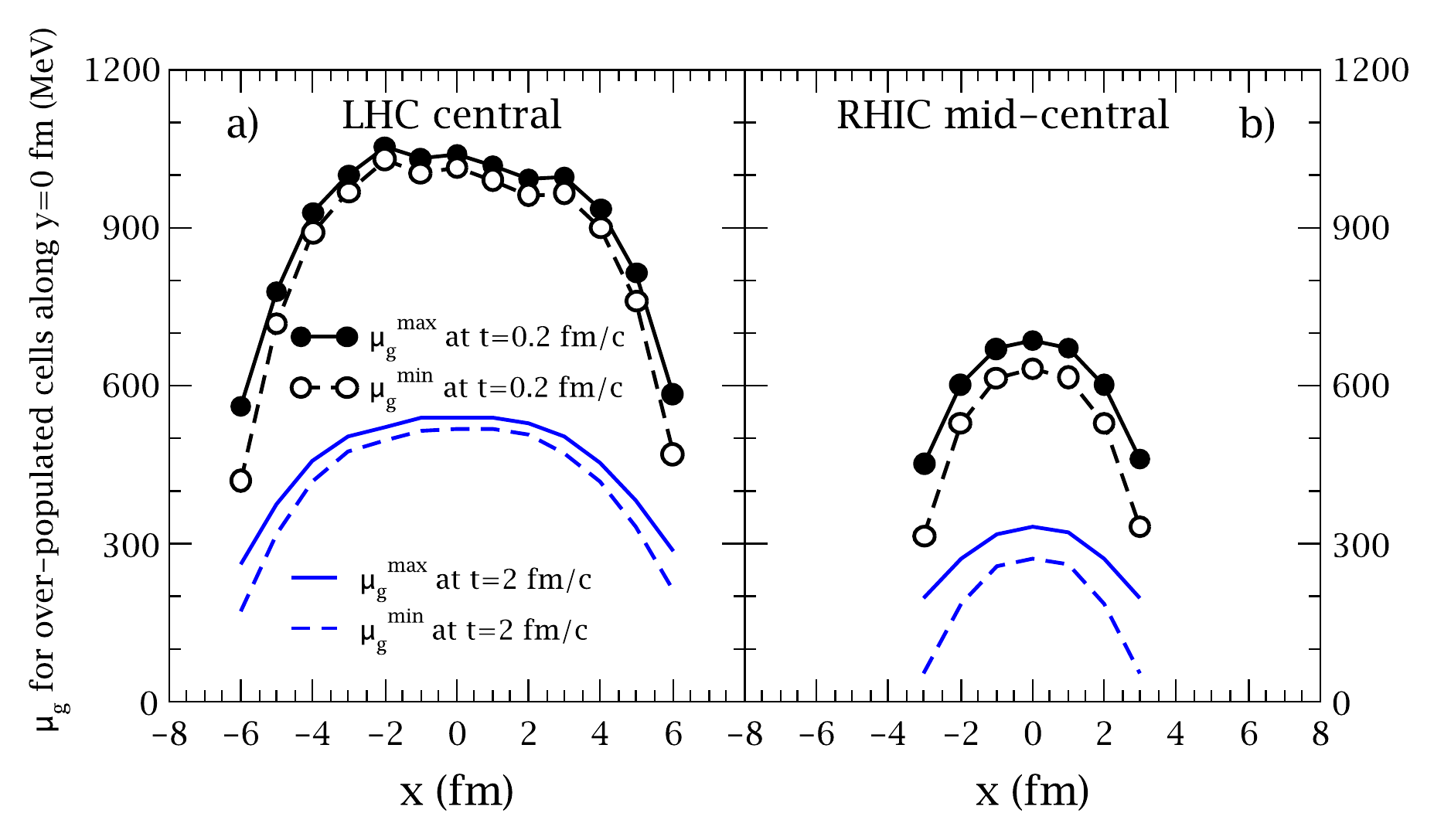}
\caption{(Color online)  
Extracted range of $\mu_g$, the gluon ``chemical potential'' parameter, 
in over-populated cells along the impact parameter at $t=0.2$ and
$2.0$ fm/$c$: 
a) for central Pb+Pb collisions at $2760A$ GeV,
and b) for mid-central Au+Au collisions at $200A$ GeV.
}
\label{fig:mug}
\end{figure}

The gluon phase-space occupancy $\gamma_g$ may be conveniently
translated into the gluon ``chemical potential'' parameter $\mu_g$ via
the definition 
\ber
\gamma_g \equiv e^{(\mu_g/T_{<p_{\rm T}>})}.
\eer
I can then obtain the range of $\mu_g$ as $\left [ \mu_g^{min},
  \mu_g^{max} \right ]$ that corresponds to 
the $\gamma_g$ range $[\gamma_g^{min},\gamma_g^{max}]$.
Fig.~\ref{fig:mug} shows the extracted range of $\mu_g$,
which corresponds to the gluon phase-space occupancy $\gamma_g$ of 
Eq.~(\ref{mug}) for the non-equilibrium quantum distribution, 
in over-populated cells along the impact parameter axis at two
different times for central Pb+Pb collisions at $2760A$ GeV 
and mid-central Au+Au collisions at $200A$ GeV. 
Although the quark and anti-quark phase-space occupancy $\gamma_q$
cannot be determined, we see that this does not lead to big
uncertainties in the extracted gluon ``chemical potential'' parameter
$\mu_g$.  We also see that the gluon ``chemical potential'' is larger
in the inner part of the overlap volume and can reach 1~GeV in the
very early stage of central Pb+Pb collisions at LHC.

\section{Discussions}

The string melting version \cite{Lin:2001zk,Lin:2004en} of the AMPT
model only has quarks and anti-quarks in the parton phase. 
This is not to be considered as physical though, since one expects gluons
to dominate the initial stage of ultra-relativistic heavy ion
collisions and gluons will also be produced from quark and anti-quark
interactions. The lack of gluons in the AMPT-SM model results from our
inability to consistently address gluon and quark productions from a
strong color field \cite{Lin:2004en}. 
However, the AMPT-SM model enables us to include the energy from all
the excited strings in the overlap volume into the parton transport.  
For studies that depend more on the effect of partonic 
scatterings instead of the composition of the partonic matter,  
such as studies of the elliptic flow \cite{Lin:2001zk} and the pion 
interferometry \cite{Lin:2002gc}, it is suitable to use the AMPT-SM
model. Since this study addresses the time evolution of parton flow,
mean momentum or energy, number density, and energy density, its
results depend mostly on the effect of partonic scatterings but not on 
the composition of the partonic matter. 
I then infer from the effective temperatures $\te$ and $\tmpt$, 
which are extracted respectively from the parton energy density and
mean $\pt$, that the high density matter in the inner region of the
overlap volume in these collisions are over-populated with partons (at
least over-populated with gluons), even though the AMPT-SM model has
no gluons in the parton phase. 
However, note that further studies, such as 
a consistent description of the initial production of gluons and
quarks from strong color fields,
will be needed to investigate whether this phase-space over-population
is a general feature \cite{Blaizot:2011xf,Berges:2012us} of the QCD matter
created in high energy heavy ion collisions. Currently I may conclude
that this is the feature of the parton system within the string
melting version of our microscopic transport model, which has been
constrained to reproduce the bulk data on the rapidity density,
$\pt$spectrum and elliptic flow at low $\pt$.  

The parton cascade in the AMPT model only includes elastic parton
scatterings. Including inelastic parton scatterings in the parton
cascade \cite{Xu:2004mz,Xu:2007aa} would change the space-time
evolution of the dense parton matter and thus affect the transverse
flow and effective temperatures. 
However, since the initial condition from the AMPT-SM model
would not be affected by the subsequent parton scatterings, the
results in this study at $t=0.2$~fm/$c$ would remain the
same. Therefore the qualitative conclusions about the mismatch among
different effective temperatures and about the over-population in the
inner part of the overlap volume would not change; while the time 
evolutions of the temperature mismatch and over-populated areas would
be affected by inelastic parton scatterings.

The space-time evolution data obtained from this study 
may serve as a bulk matter background for studies such as jet
propagation and interactions with the partonic matter within the JET
Collaboration \cite{Burke:2013yra,jetcollab} and beyond.  
The effective temperatures extracted from the parton phase in
the AMPT-SM model also provide a link to hydrodynamic models. 
Currently one may argue that $\te$ is a reasonable choice for the effective
temperature of the parton system, because it takes into account both
the density and the mean momentum of the local parton system.   
Also, the essential variable in the hydrodynamic approach is the local
energy-momentum tensor, and using $\te$ ensures that the local
energy distributions remain the same when the transport approach is
linked to the hydrodynamic approach. 
However, since the effective temperatures depend on the variables from
which they are extracted, further studies will be necessary to address the
uncertainty associated with this dependence. As we have seen, the
mismatch between $\te$ and $\tmpt$ could be attributed to
an over-population of partons in the dense matter, thus the
uncertainty about the effective temperatures reflects the uncertainty
in the initial conditions for ultra-relativistic heavy ion collisions.

\section{Conclusions}

I have studied the space-time evolution of the parton phase created
in heavy ion collisions at RHIC and LHC energies using the AMPT
model with string melting, which converts excited hadronic strings in
the overlap volume into partons. Several key parameters in the model
have been tuned to reproduce the low-$\pt$ data on the pion and kaon
yields, $\pt$ spectra and elliptic flows. This way
the space-time evolution from the model is more reliable, and it may
serve as a bulk matter background for studies such as jet
propagation and interactions with the partonic matter.

This study focuses on the effective temperatures in different volume
cells within mid-spacetime-rapidity $|\eta|<1/2$. Many events at
the same impact parameter are averaged over in order to have enough 
statistics for the analysis of each volume cell, and as a result this study
does not address the effect of event-by-event fluctuations. 
Since the parton system in a subvolume is in general
neither in full thermal equilibrium nor in full chemical equilibrium,    
the value of the effective temperature depends, sometimes strongly, on
the variable that it is extracted from. 
I have extracted effective temperatures from several different
variables that are evaluated in the rest frame of each cell. 
I find that temperature $\te$ extracted from 
the parton energy density is often very different from 
$\tmpt$ extracted from the parton mean $\pt$,  
and it is mostly between $\tmpt$ and $\tn$ extracted from the parton
number density while being closer to $\tn$.
For these collisions I also find that $\te > \tmpt$ over the inner
part of the overlap volume, which indicates that partons (at least 
gluons) are over-populated there. This is also checked by 
examinations of parton phase-space densities, 
parton phase-space occupancy factors, and the gluon ``chemical
potential'' parameter.
My results show that at mid-spacetime-rapidity initially about half
or more QGP cells, defined as cells that are above a critical energy
density of $1$ GeV/fm$^3$, are over-populated. 
The initial fraction of over-populated cells is found to be bigger 
at higher energies and in more central collisions. 
I also find that all QGP cells are over-populated after a couple
of fm/$c$, and the total transverse area of over-populated cells
does not change much during the first few fm/$c$.

\section{Acknowledgments}
The author would like to thank B. Betz for valuable discussions and express 
special thanks to M. Gyulassy for motivating this study. 
As an external associate, the author thanks the JET Collaboration
for help. The author also thanks Dr. X.-N. Wang for updating the JET
Collaboration wiki page with the link to the data files from this study.


\begin{thebibliography}{99}

\bibitem{Alver:2010gr} 
  B.~Alver and G.~Roland,
  Phys.\ Rev.\ C {\bf 81}, 054905 (2010)
  [Erratum-ibid.\ C {\bf 82}, 039903 (2010)].

\bibitem{Blaizot:2011xf} 
  J.~-P.~Blaizot, F.~Gelis, J.~-F.~Liao, L.~McLerran and R.~Venugopalan,
  Nucl.\ Phys.\ A {\bf 873}, 68 (2012).

\bibitem{Betz:2012qq} 
  B.~Betz and M.~Gyulassy,
  Phys.\ Rev.\ C {\bf 86}, 024903 (2012).

\bibitem{CMS:2012qk} 
  S.~Chatrchyan {\it et al.}  [CMS Collaboration],
  Phys.\ Lett.\ B {\bf 718}, 795 (2013).

\bibitem{Huovinen:2001cy} 
  P.~Huovinen, P.~F.~Kolb, U.~W.~Heinz, P.~V.~Ruuskanen and S.~A.~Voloshin,
  Phys.\ Lett.\ B {\bf 503}, 58 (2001).

\bibitem{Betz:2008ka} 
  B.~Betz, J.~Noronha, G.~Torrieri, M.~Gyulassy, I.~Mishustin and D.~H.~Rischke,
  Phys.\ Rev.\ C {\bf 79}, 034902 (2009).

\bibitem{Schenke:2010rr} 
  B.~Schenke, S.~Jeon and C.~Gale,
  Phys.\ Rev.\ Lett.\  {\bf 106}, 042301 (2011).

\bibitem{Bozek:2011if}
  P.~Bozek,
  Phys.\ Rev.\ C {\bf 85} (2012) 014911.

\bibitem{Xu:2004mz} 
  Z.~Xu and C.~Greiner,
  Phys.\ Rev.\ C {\bf 71}, 064901 (2005).

\bibitem{Lin:2004en} 
  Z.~-W.~Lin, C.~M.~Ko, B.~-A.~Li, B.~Zhang and S.~Pal,
  Phys.\ Rev.\ C {\bf 72}, 064901 (2005).

\bibitem{Cassing:2009vt} 
  W.~Cassing and E.~L.~Bratkovskaya,
  Nucl.\ Phys.\ A {\bf 831}, 215 (2009).

\bibitem{Petersen:2008dd} 
  H.~Petersen, J.~Steinheimer, G.~Burau, M.~Bleicher and H.~Stocker,
  Phys.\ Rev.\ C {\bf 78}, 044901 (2008).

\bibitem{Werner:2010aa} 
  K.~Werner, I.~.Karpenko, T.~Pierog, M.~Bleicher and K.~Mikhailov,
  Phys.\ Rev.\ C {\bf 82}, 044904 (2010).

\bibitem{Song:2010mg} 
  H.~Song, S.~A.~Bass, U.~Heinz, T.~Hirano and C.~Shen,
  Phys.\ Rev.\ Lett.\  {\bf 106}, 192301 (2011)
  [Erratum-ibid.\  {\bf 109}, 139904 (2012)].

\bibitem{Lin:2001zk} 
  Z.~-W.~Lin and C.~M.~Ko,
  Phys.\ Rev.\ C {\bf 65}, 034904 (2002).

\bibitem{Zhang:1999bd} 
  B.~Zhang, C.~M.~Ko, B.~-A.~Li and Z.~-W.~Lin,
  Phys.\ Rev.\ C {\bf 61}, 067901 (2000).

\bibitem{Lin:2000cx} 
  Z.~-W.~Lin, S.~Pal, C.~M.~Ko, B.~-A.~Li and B.~Zhang,
  Phys.\ Rev.\ C {\bf 64}, 011902 (2001).

\bibitem{Sjostrand:1993yb} 
  T.~Sjostrand,
  Comput.\ Phys.\ Commun.\  {\bf 82}, 74 (1994).

\bibitem{Xu:2011fi} 
  J.~Xu and C.~M.~Ko,
  Phys.\ Rev.\ C {\bf 83}, 034904 (2011).

\bibitem{datalink} 
Evolution data files at http://myweb.ecu.edu/linz/ampt/evolutiondata/.

\bibitem{jetwiki} 
The JET Collaboration wiki page at 
https://sites.google.com/a/lbl.gov/jetwiki/code-packages/hydro-evolution/.

\bibitem{Betz:2013caa} 
  B.~Betz and M.~Gyulassy,
  arXiv:1305.6458 [nucl-th].

\bibitem{Burke:2013yra} 
  K.~M.~Burke, A.~Buzzatti, N.~Chang, C.~Gale, M.~Gyulassy, U.~Heinz, S.~Jeon and A.~Majumder {\it et al.},
  arXiv:1312.5003 [nucl-th].

\bibitem{Abelev:2013vea} 
  B.~Abelev {\it et al.}  [ALICE Collaboration],
  Phys.\ Rev.\ C {\bf 88}, 044910 (2013).

\bibitem{Adare:2012wg} 
  A.~Adare {\it et al.}  [PHENIX Collaboration],
  Phys.\ Rev.\ C {\bf 87}, 034911 (2013).

\bibitem{Abelev:2013qoq} 
  B.~Abelev {\it et al.}  [ALICE Collaboration],
  Phys.\ Rev.\ C {\bf 88}, 044909 (2013).

\bibitem{Adler:2003cb} 
  S.~S.~Adler {\it et al.}  [PHENIX Collaboration],
  Phys.\ Rev.\ C {\bf 69}, 034909 (2004).

\bibitem{Bearden:2004yx} 
  I.~G.~Bearden {\it et al.}  [BRAHMS Collaboration],
  Phys.\ Rev.\ Lett.\  {\bf 94}, 162301 (2005).

\bibitem{Gu:2012br} 
  Y.~Gu [PHENIX Collaboration],
  Nucl.\ Phys.\ A {\bf 904-905}, 353c (2013).

\bibitem{ATLAS:2012at} 
  G.~Aad {\it et al.}  [ATLAS Collaboration],
  Phys.\ Rev.\ C {\bf 86}, 014907 (2012).

\bibitem{Zhang:2008zzk} 
  B.~Zhang, L.~-W.~Chen and C.~M.~Ko,
  J.\ Phys.\ G {\bf 35}, 065103 (2008).

\bibitem{Florkowski:2010cf} 
  W.~Florkowski and R.~Ryblewski,
  Phys.\ Rev.\ C {\bf 83}, 034907 (2011).

\bibitem{Martinez:2010sc} 
  M.~Martinez and M.~Strickland,
  Nucl.\ Phys.\ A {\bf 848}, 183 (2010).

\bibitem{Bazow:2013ifa} 
  D.~Bazow, U.~W.~Heinz and M.~Strickland,
  arXiv:1311.6720 [nucl-th].

\bibitem{Xu:2007aa} 
  Z.~Xu and C.~Greiner,
  Phys.\ Rev.\ C {\bf 76}, 024911 (2007).

\bibitem{Lappi:2006fp} 
  T.~Lappi and L.~McLerran,
  Nucl.\ Phys.\ A {\bf 772}, 200 (2006).

\bibitem{Lin:2002gc}
Z.~W.~Lin, C.~M.~Ko and S.~Pal,
Phys.\ Rev.\ Lett.\  {\bf 89}, 152301 (2002).

\bibitem{Berges:2012us} 
J.~Berges and D.~Sexty,
Phys.\ Rev.\ Lett.\  {\bf 108}, 161601 (2012).

\bibitem{jetcollab}
Topical Collaboration on Jet and Electromagnetic Tomography of Extreme
Phases of Matter in Heavy-ion Collisions at http://jet.lbl.gov/.

\end{thebibliography}
\end{document}